\documentstyle[12pt,aaspp4]{article}

\begin{document}

\def\etal{{\it et al.\/\ }}
\def\today{\ifcase\month
           \or January   \or February \or March    \or April 
           \or May       \or June     \or July     \or August
           \or September \or October  \or November \or December\fi
           \space\number\day, \number\year}
\def\ga{\lower 2pt \hbox{$\, \buildrel {\scriptstyle >}\over{\scriptstyle \sim}\,$}}
\def\la{\lower 2pt \hbox{$\, \buildrel {\scriptstyle <}\over{\scriptstyle \sim}\,$}}

\title{\bf EXPANDING THE REALM OF MICROLENSING SURVEYS WITH 
DIFFERENCE IMAGE PHOTOMETRY}

\author{Austin B. Tomaney$^{1,2}$ and Arlin P. S. Crotts$^{1,2}$}
\affil{Department of Astronomy, Columbia University, 538 W.~120th St., 
New York, NY~~10027}
\affil{\bigskip
$^1$ Based on observations at the 
Kitt Peak National Observatory, National Optical
Astronomy Observatories, operated by the Association of Universities for
Research in Astronomy, Inc., under contract with the National Science
Foundation
}

\affil{\bigskip
$^2$ Based on observations at the
Vatican Advanced Technology Telescope (The Alice P. Lennon Telescope
and Thomas J. Bannan Astrophysical Facility), Mt Graham Arizona
}

\authoremail{austin@odyssey.phys.columbia.edu, arlin@eureka.phys.columbia.edu}

\begin{abstract}
We present a new technique for monitoring microlensing activity even in highly
crowded fields, and use this technique to place limits on low-mass MACHOs in
the haloes of M31 and the Galaxy.
Unlike present Galactic microlensing surveys, we employ a technique in which a
large fraction of the stellar sample is compressed into a single CCD field,
rather than spread out in a way requiring many different telescope pointings.
We implement the suggestion by Crotts (1992) that crowded fields can be
monitored by searching for changes in flux of variable objects by subtracting
images of the same field, taken in time sequence, positionally registered,
photometrically normalized, then subtracted from one another (or a sequence
average).
The present work tackles the most difficult part of this task, the adjustment
of the point spread function among images in the sequence so that seeing
variations play an insignificant role in determining the residual after
subtraction.
The interesting signal following this process consists of positive and negative
point sources due to variable sources.
The measurement of changes in flux determined in this way we dub ``difference
image photometry'' (also called ``pixel lensing'' [Gould 1996]).

The matching of the image point spread function (PSF) is accomplished by a
division of PSFs in Fourier space to produce a convolution kernel, in a manner
explored for other reasons by Phillips \& Davis (1995).
In practice, we find the application of this method is difficult in a typical 
telescope and wide field imaging camera due to a subtle interplay between 
the spatial variation of the PSF associated with the optical design and
the inevitable time variability of the telescope focus. Such effects lead to
complexities in matching the PSF over an entire frame.
We demonstrate the realization of the difference image approach with two
separate 
solutions to these problems - a software algorithm to determine the match
of the spatially varying PSF between frames using a limited number of stars 
and also a simple optical corrector for a wide field imager to simplify the
PSF matching function.

The former solution yielded light curves of 139 variable
sources detected in a 16$'$ by 16$'$ field in M31 over four nights on the
KPNO 4-m telescope in 1994 and the latter yielded
over 2000 sources detected over 50 nights
in a single 11$'$ by 11$'$ field in M31 observed at the VATT 1.8-m 
telescope in 1995 using an optical corrector to facilitate the PSF
matching problem. Of the KPNO sources discovered, 85 overlap with the 
VATT field and 23 of these were found to have a positional coincidence 
of $< 1''$ to sources found in the VATT data. Light curves of the 
VATT objects over 14 nights confirm the short timescale
variability of these sources.
Although some fraction of the sources are bright enough to be considered
resolved in the raw data more than half the sources are fainter than
the surface brightness fluctuations associated with the unresolved 
stars in the galaxy and cannot be identified in the raw data. However,
the light curves of these sources appear to be familiar variables such
as Cepheids and eclipsing binaries.

We assess the limitations and sensitivities of the techniques and demonstrate
that we can achieve photometric errors of faint unresolved variables that
are within a factor of three of the ultimate photon noise limit. Using this
we show how the KPNO data over two good nights, and sensitive to $> 10^6$
stars on a single CCD frame, 
yields $2\sigma$ optical depth limits of $5\times 10^{-7}$
for Galactic MACHOs in the mass range $2\times 10^{-7} {\rm M_\odot}$ 
($0.07$ Earth masses). Given the estimate of
the optical depth of the Galactic halo towards M31 of $\tau = 1\times 10^{-6}$
(assuming a simple spherical halo), 
we can conclude that in two nights we have eliminated 
the possibility at the $95$\% confidence level 
that the Galactic halo is comprised of a single mass population
of MACHOs in the sub-Earth mass range. Based on estimates of the M31 and 
Galactic MACHO $\tau = (5-10)\times 10^{-6}$ we exclude the halo of these
galaxies being composed of $8\times 10^{-5} {\rm M_\odot}$ MACHOs at the 
$> 95\%$ confidence level.

These kind of techniques can extend present microlensing surveys 
into regimes not limited to resolved stars,
which greatly expands the power of these surveys.
Application to surveys of more general kinds of variability is clear.
We also suggest other applications, such as to proper motion surveys.
\end{abstract}

\clearpage

\section{INTRODUCTION}

The potential rewards of a microlensing survey of M31 were first outlined
by Crotts (1992, hereafter C92) and also analyzed in other papers by 
Baillon \etal (1993) and Colley (1995). In addition to Galactic halo MACHOs
(MAssive Compact Halo Objects) such a survey is also sensitive to MACHOs
in the halo of M31. Specifically, for M31 MACHOs large optical depths,
$\tau$ (lensing probability per star), 
can be achieved considerably in excess of equivalent Galactic surveys,
since a line of sight can be chosen through the densest regions of the
halo.
Due to the high inclination of M31, and depending on the halo and bulge
geometry,
optical depths can be up to ten times higher than for Galactic MACHOs.
Since the lensing optical depth is dependent on
the halo geometry, measurement of the optical depth can constrain halo
models (see also Gould 1994 and Han \& Gould 1996). 
This modulation can also allow for the construction
of a control experiment in which equivalent stellar populations at the
same galactocentric distance are monitored but with different lensing
rates, in particular when comparing the near and far side of the 
galaxy.

An important additional benefit occurs for Galactic MACHOs. For a
high amplification
event to be detected, the projected Einstein ring at the lensed star must
exceed the size of the photosphere of the star. Since the Galactic lenses
are much closer they can be lower mass and still produce a
detectable event. Thus sensitivity is dramatically increased for much
lower masses in this case.

Unfortunately, at the distance of M31 (770 kpc, Freedman \& Madore
1990) almost all
stars are completely unresolved in typical ground based seeing of 
$\sim 1''$. Given present estimates of the optical depth towards M31 of up 
to $10^{-5}$ for the combination of both Galactic and M31 MACHOs (C92, 
Han \& Gould 1996)
an observing program must be sensitive to at least 
$10^5$ to $10^6$ stars to have a good probability of detecting a lensing 
event. It appears to be extremely unlikely in a given field of stars in 
M31 that a lensing event will occur in a bright star that could be 
considered to be resolved from the ground with present imaging capabilities.

To cope with this problem C92 suggested registering a sequence of CCD
images to common coordinates, scaling to the same photometric intensity
and subtracting images from a high S/N (signal to noise) template image. 
Since over some timescales of interest for microlensing events
($10^m$ up to a few months), {\it most} stars will not be varying (at least
above some given detection limit in flux change),
then those stars that do vary over the 
time span between the test frame and the reference frame may be detected 
in the difference frame as isolated positive or negative point sources 
depending on whether the particular star brightened or faded with respect
to the reference frame.
We call this ``Difference Image Photometry'' (DIP).
Such a technique has recently been dubbed ``pixel
lensing'' and has been theoretically formalized in Gould (1996). 

Although frame registration and photometric scaling are quite tractable 
with presently available software such as IRAF, it is the seeing variations
between frames that cause the most concern for this technique. Frames 
not well matched in seeing will result in power on the scale of the PSF 
in the difference frame as flux from unresolved stars is poorly subtracted.
Such systematic residuals may swamp the signal of genuine variability.
It is this issue that in the past 
caused skepticism as to whether this kind of 
microlensing survey could ever be realised from the ground. 

We employ a Fourier technique first outlined 
in Ciardullo, Tamblyn \& Phillips 
(1990) and further developed by Phillips \& Davis (1995)
to match the PSF between frames. In practice, we find that the application 
of the technique to data taken with the KPNO 4-m prime focus camera is not
a complete solution to PSF matching, and find that, separate from seeing
variations, {\it focus} variations of the telescope increase the complexity
of the PSF matching over the entire CCD frame. We outline an algorithm
we have developed to cope with this problem which uses a limited number of
stars on a given frame to determine the correct full-frame PSF matching 
function 
for a pair of frames to be differenced. We present the results from the
application of this algorithm to four nights of KPNO 4-m data here.

In the light of the challenges posed by the KPNO data, we discuss how PSF
matching can be facilitated with a simple optical corrector in a wide field
imager. We demonstrate such a solution with a camera we have designed for
the VATT 1.8-m telescope. Some preliminary results of an analysis of a 
subset of data taken at the VATT in 1995 as part of an ongoing survey 
to discover microlensing in M31 are also presented.

The outline of the paper is as follows. In {\S 2} we discuss the observational
aspects surveys geared to the discovery of microlensing in M31 and the details
of the observations we have carried out at KPNO to discover microlensing 
by Galactic MACHOs around $10^{-6} {\rm M_\odot}$ range, and 
more generally M31 MACHOs with masses of 
$\sim 10^{-4} {\rm M_\odot}$ and above using the VATT 1.8-m
telescope. In {\S 3} we discuss the difference image technique, 
including the algorithms we use to
match the PSF between frames. 
In {\S 4} we provide a detailed discussion of the important systematic
effects involved in image differencing and
assess the technique photometric
sensitivities to faint, unresolved stars where we 
demonstrate that we can achieve photometric errors that are within a 
factor of three or less of the ultimate photon noise limit. 
In {\S 5} we present results from a difference image
analysis of the KPNO data and a similar
preliminary analysis of the VATT data. We show light curves of variables
discovered independently in both data sets including many Cepheid and 
eclipsing binary candidates. 
In {\S 6} we assess the sensitivity of the KPNO data
in terms of the number of stars, microlensing timescales and MACHO masses 
and the search for microlensing on the 50$^m$ to 8$^h$ timescales 
associated with Galactic MACHOs in the mass range $10^{-7}$ to $10^{-5}
{\rm M_\odot}$ and M31 and Galactic MACHOs of 
$\sim 10^{-4} {\rm M_\odot}$. 
We outline an estimate of optical depth limits we
have achieved in this range.
In {\S 7} we summarize our conclusions and suggest other possible
microlensing related applications of DIP.

\section{THE SURVEYS}

We outline the broad aspects of our surveys here and 
discuss in detail the number of stars we are
sensitive to, in addition to timescale and mass sensitivities in {\S 6}.

\subsection{Target Stars}

Typically the brightest stars in M31's bulge and inner disk
region are red giants (RGs).
A microlensing survey needs to maximize detector sensitivity to these 
objects, preferably in two bandpasses to test for achromaticity of
light curves which must not exhibit color variations (Paczynski 1986, but
see also Kamionkowski 1995). We choose non-standard R and I filters, 
slightly broader than their conventional equivalents: R extends from 
$\lambda5700$ (just beyond the [O I] $\lambda5577$ night sky line) to
$\lambda7100$, and filter I extends from $\lambda7300$ to $\lambda10300$
(5\% power points).
These choices maximize the number of photons in each filter we can detect
from RGs with a CCD.

\subsection{The KPNO Survey: Testing the Galactic MACHO $10^{-7} {\rm M_\odot}$
regime}.
 
We observed four nights (1994 September 24-27, UT)
on the KPNO 4-m with the $16' \times 16'$
field of view prime focus camera - a Tek $2048^2$ CCD 
(TK2B) with a platescale of $0.48''$ per pixel.
The first two nights were plagued by
clouds, moonlight and poor seeing. The last two nights were predominately
photometric with $\sim 1.1-2.2''$ seeing. The target field was centered 
in the maximal lensing field (MLF) predicted in C92 to be at about 1.5 kpc
(7.5$'$) from the nucleus along the minor axis on the far side of the disk
for a simple halo models ($\rho \propto r^{-2}$).
This is also the high $\tau$ region from later models (Han \& Gould 1996).
This field includes the nucleus, which typically saturates very quickly in a
given exposure.
Exposure times were limited to 150s so that no more than a
$2.5' \times 1.5'$ region centered on the nucleus was saturated on the CCD.
The analysis presented here is based on coadded exposures totaling
$12.5^m$ in both R and I filters. Given the CCD readout overhead for a
large CCD, this corresponds to a time resolution of $50^m$ per coadded frame
in a given band
(about four times the minimum-mass Einstein crossing time, $t_{_E}$, which is
just $R_E$ divided by the transverse velocity of the lens with respect to the
observer-source sightline).
Such time resolution yields sensitivity to Galactic MACHOs around 
$2\times10^{-7}
{\rm M_\odot}$ given the assumed stellar sizes we outline in {\S 6.2} 
({\it c.f.} C92).
 
From our photometric calibrations ({\S 5}) and
integrating the {\it unsaturated} regions of a typical coadded exposure
we obtain
a mean surface brightness of galaxy plus sky of $\mu_R = 19.35$ 
and $\mu_I = 18.30$.
This corresponds to a mean $\langle S/N \rangle$ of 16 for an R = 22.5 star in
the frame.
Above this magnitude cutoff we are sensitive to $6.7\times10^5$ stars at
greater $\langle S/N \rangle$ in this single field using our
number density estimate in {\S 6.1}.
 
\noindent
\subsection{The VATT Survey: Probing the Halo of M31 for MACHOs}
 
We obtained 26 nights of data spread over a 50 night timespan on the
new VATT 1.8-m telescope in the Fall of 1995 with an $11.3'\times11.3'$
field of view camera with a Tek$2048^2$ CCD with a platescale of
$0.33''$ per pixel. The typical seeing range
in these data was $0.8''$ to $1.5''$ with a median seeing around $1.0''$.
The field observed included the MLF field and its near side equivalent,
which is predicted to have an optical depth up to an order of magnitude
lower than the MLF depending on the halo geometry.
The MLF field was observed in both R and I.
Under typical photometric, moonless arcsecond observing conditions
a $60^m$ integration in R corresponds to a $\langle S/N \rangle$ of 16 for an
R=22.5 star in these fields.
This magnitude limit corresponds to $\sim 3.1\times 10^5$ stars ({\S 6.1}). 
Exposures were adjusted under varying conditions to ensure
that equivalent depths were achieved in at least the MLF field in R
on any given night.
Further coverage for the 1995/1996 observing season was also obtained with
the Wise Observatory 1.0-m telescope; extensive results from these
datasets will be presented in future papers.
 
\section{DIFFERENCE IMAGE PHOTOMETRY}

The following discussion is confined to the KPNO dataset.
We use the VATT data in this paper to illustrate a simplifying approach to 
matching the PSF between frames ({\S 3.6}), 
as well as independently confirm the 
reality of some of the variable sources detected in the KPNO data ({\S 5}).
VATT data were reduced along similar lines to the KPNO data 
and a more detailed 
discussion of the processing will be presented in a future paper.

\subsection{Preliminary Processing}

All frames were debiased and flat-fielded in the standard manner. Sky-flats
were used to divide out the overall illumination of the CCD in each filter
and dome flats used to derive the pixel-to-pixel variations.

Each frame was cleaned individually of bad pixels including cosmic rays. 
Such pixels were identified by fitting a 5x5 pixel
2D Gaussian of width less than the
seeing on each frame to each pixel in the image. The residuals of
the difference between the data and the model effectively discriminate 
between neighboring pixels that are consistent with the seeing and those
that are associated with chip defects and cosmic rays. Once such pixels 
were identified simple linear interpolation using good neighboring
pixels was used to replace a bad pixel. The advantage of cleaning individual
frames in this way is to allow a weighted combination of frames taken 
under variable seeing to be made which is clean of defects, 
as compared with a more robust combination
such as a median which can degrade the final PSF of a
combination of frames with very different seeing. 

Registration of all frames to common coordinates was made with $\sim 50$
bright, unsaturated stars on each frame whose centers were determined 
using the DAOPHOT PHOT routine and the IRAF routines GEOMAP and GEOTRAN. 
A 5th order polynomial interpolant was used in the geometric transformation
and flux was conserved. 
Final registration of frames was accurate to within 0.1 pixels. 
The PSFs of stars on all frames were well sampled with minimum FWHM's
of $2.1$ pixels, thus resampling errors were small and PSFs were not
significantly degraded in this registration process.
At this point frames were combined for different timescales:
for the analysis of sub-night timescales sequences typically five 
images were combined 
giving images with exposure times of 12.5$^m$ in each filter, 
and on nightly timescales all frames taken in each filter were
combined.

\subsection{PSF Matching with a Fourier Algorithm}

In some test frames taken in similar seeing C92 was able to show 
that the residuals
in the difference image were comparable to the photon noise. 
As stressed in {\S 1}, however, the residuals in the difference image are
expected to be primarily influenced by differences in seeing between
a pair of frames to be differenced. Since a typical PSF approximates 
a Gaussian we have found that simply convolving one frame with a Gaussian
kernel to match the FWHM of the PSF of the better seeing frame is very
effective. In general, systematic residuals in the difference frame can
be minimized by matching the PSFs between frames as closely as possible.

In order to monitor partially resolved globular clusters in M31 
for nova eruptions over a number of years through differential
photometry Ciardullo, Tamblyn \& Phillips (1990) 
developed a Fourier algorithm 
to match seeing variations of different epochs frames. In essence
each good seeing frame was degraded to a common seeing value.
With frames of identical seeing
the relative flux of all non-varying point-like sources in a 
frame is the same for any given finite photometric aperture and thus
meaningful differential photometry can be performed between frames.
Phillips \& Davis (1995) have developed this algorithm further and
we are very grateful to Andrew Phillips for providing us with 
his software, which we have employed as
the basis of our PSF matching technique. To apply
it to its full extent to a typical wide field imaging telescope is
not straightforward. In {\S 3.3} we discuss the details and the 
algorithm we have developed to solve these problems and apply this
to the KPNO data. In {\S 3.6} we
show how the PSF matching can be simplified with an optical corrector
for a wide field imager and demonstrate its application to
the VATT data.

First, ignoring noise,
consider a frame $r$ with a narrower PSF than a frame $i$,

$$ i = r * k, \eqno (1) $$

\noindent
where $k$ is a convolution kernel that describes the difference in the 
seeing and guiding between the two frames. The Convolution Theorem
states that the Fourier transform (FT) of these three variables
has the form,

$$ FT (i) = FT (r) \times FT (k), \eqno (2) $$

\noindent
then,

$$ k = FT [ FT (i) / FT (r) ], \eqno (3) $$

\noindent
Thus $k$ can be determined empirically with a high S/N isolated 
star on a frame pair. Convolving the good seeing frame with this
kernel will in principle provide a match to the PSF of the poorer
seeing frame.
In practice the determination of $k$ is not straightforward 
since the high frequency components of the FT become dominated by the 
noise in the wings of the PSF where the signal is weakest. An effective
method of dealing with this problem was determined by Ciardullo,
Tamblyn \& Phillips
(1990). Since the FT of a typical PSF is roughly Gaussian the convolution 
kernel will also be approximately Gaussian. By modeling the high
S/N low-frequency components of the PSF FT with an elliptical 
Gaussian these noise-contaminated components can be replaced with the
model fit yielding a convolution kernel close to the ideal. 

Figure 1 shows an application of this method to some KPNO data. We
show a 128x128 pixel subimage near the center of an image,  
which clearly shows a background of unresolved stars. 
A suitable bright star near this region has been used to 
determine $k$. Typically the kernel is determined over a region 
extending up to five times the FWHM of the PSF. After photometric
scaling of the frame pair ({\S 3.5}) and the degradation of the better seeing
frame with its convolution with $k$, the difference image
shows the clear removal of structure in the background, including
a good subtraction of all but the brightest star on the frame which
is saturated. However, also shown in Figure 1 is a subimage of the 
difference frame located about
500 pixels away from the location where $k$ was determined. It is clear
that this region is plagued by large systematic residuals associated 
with a poorly matched PSF in this region. Critically, this is not
simply a problem affecting the bright stars on the frame, but also 
the background of unresolved stars where we are most concerned with 
achieving the best subtraction. This demonstrates that there is no unique 
solution of $k$ applicable to the entire frame.

\begin{figure}
\plotfiddle{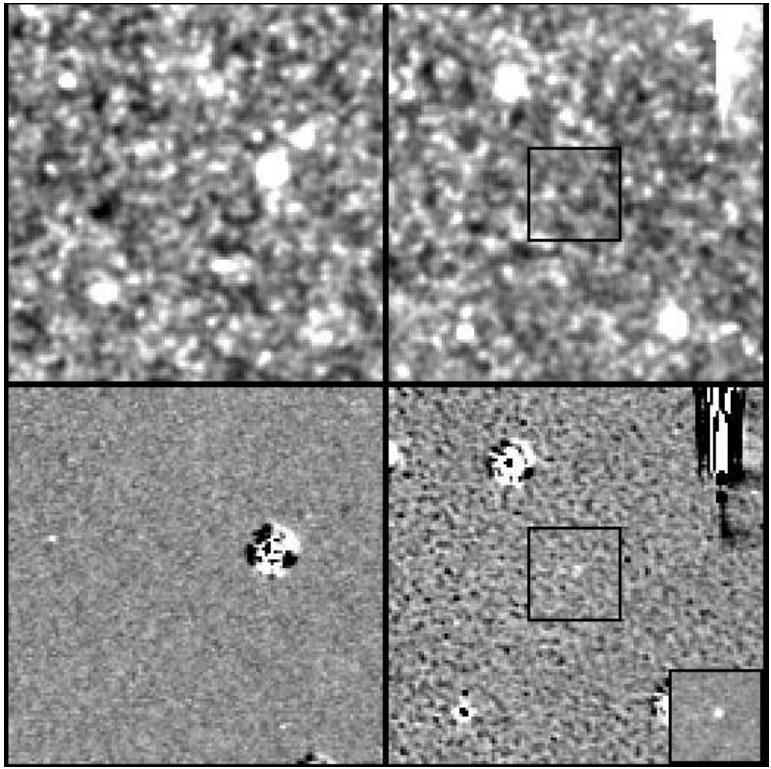}{4in}{0.0}{50}{50}{-288}{-648}
\caption {
Application of the Phillips \& Davis (1995) algorithm
to the KPNO data. The left side shows a $128\times 128$ pixel
subimage close to the center of the original frame
(upper left panel) with the mean galaxy background
subtracted from the image, and its difference image (lower left panel).
A suitable star close to this region
has been used to empirically determine the PSF matching
convolution kernel to match the image pair being differenced
({\S 3.2}). All structure
in the unsubtracted data has been effectively removed; the residuals
around the brightest star are due to it being saturated on the CCD.
However, applying the same convolution kernel to a region located
500 pixels away (upper right panel) shows large systematic residuals
in its
difference frame on the scale of the PSF (lower right panel). This
indicates a poor match of the PSF in this region and shows that there
is no unique solution to the matching convolution kernel applicable to
the entire frame. Effective subtraction of the full image can only be
done by modeling the spatially varying PSF kernel using the limited
number of appropriate PSF matching stars on the frame {\S 3.4}.
Once applied in this case the quality of the subtraction for the entire
frame becomes comparable to the lower left panel. The inset image
in the lower right corner is the new difference image located in the
region of the box in the upper right panel. A clear detection of a
point source is now made, which was almost
completely hidden in the systematic
residuals in the first attempt at matching the PSF.
All differences are displayed in the same intensity range; the intensity
range of the upper panels is five times larger.
\label{fig1}}
\end{figure}

\subsection{Understanding the Spatial Dependence of the Matching PSF Function}

Figure 2 illustrates why $k$ becomes spatially dependent in a frame pair
to be differenced. We plot the FWHM of bright, unsaturated stars (we ignore
partially resolved M31 globular clusters) as 
a function of distance from some radial symmetry point of the PSF variation
close to the center of the CCD ({\S 3.5} shows how this was located)
for two frames taken close together in 
time and in the same filter. The radial variation of the PSF is clear
and is quite dramatic at furthest radial distance which is located near one 
corner of the CCD. However, also apparent is the different form this
radial variability takes in the two separate frames. This form is affected
by small differences in the focus of the telescope between the two exposures.
It is this time variability of the spatial functional form of the PSF that, 
independent of seeing and guiding variations which
are themselves not spatially dependent across the camera, causes the 
PSF matching convolution kernel $k$ to become a function of position on
the frame. 

\begin{figure}
\plotone{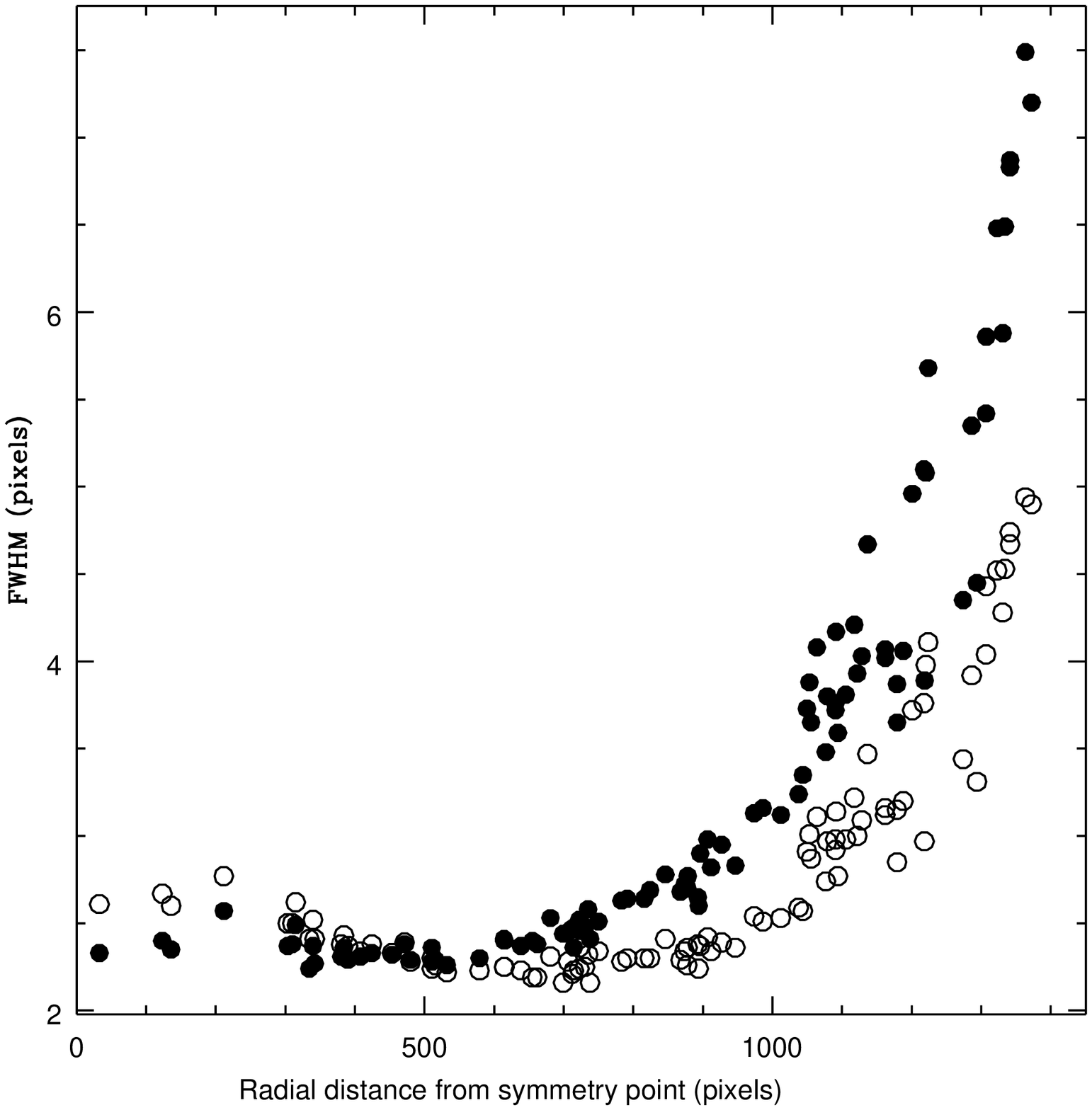}
\caption {
The FWHM of point sources on two separate R band KPNO
frames taken two hours apart (shown as open and filled circle points)
as a function of distance from the PSF radial-variation
symmetry point on each frame. The functional form of the radial variability
of the FWHM is different in the two images due to a small change in
the telescope focus between the two exposures.
\label{fig2}}
\end{figure}

In principle the solution to this problem is simply to match frames 
in a piecemeal fashion by PSF-matching subregions 
around bright, isolated and unsaturated stars which can be used to
determine the correct convolution kernel for the local region. 
In practice, however, the extent over which good subtraction can be 
made local to a PSF matching star 
is frequently only as large as $50\times 50$ pixels. Such a method 
therefore requires a star suitable for determining a convolution 
kernel in over 1500 subframes to adequately match the entire $2048^2$
pixel frame. The requirements for the choice of PSF matching stars
are rather stringent: the star must (i) have high S/N, (ii) must not
be saturated in any part of the PSF or corrupted by any defects such
as cosmic rays or bad pixels, (iii) must have an amplitude that
significantly exceeds the surface brightness fluctuations associated 
with the unresolved stars in the galaxy, and (iv) must be isolated 
from any bright neighbors over the extent to which $k$ is being 
determined. In these data the number of point-sources on each
frame that satisfy these criteria is typically $\sim 200$. Furthermore,
few of these sources reside in the bulge region where the underlying
bulge light requires correspondingly brighter stars for good PSF 
matching, leaving large and important areas of the frame without 
good kernel determinations. In the next section we describe how
this problem was addressed by deriving a model for the spatial 
variability of the matching convolution kernel over the entire
frame from every available location where local convolution kernels 
could be measured.

\subsection{Derivation of Spatially Dependent PSF-matching Convolution Kernel}

The search for variables in various frames proceeded by differencing
each frame with a high S/N template reference image with good seeing. 
For this purpose the KPNO reference images for both filters were
generated from the
combined image stacks on the final night since this
was the best quality data. 
The analysis of each frame began with measures of the raw convolution 
kernels for the individual frame/reference frame pair,
using the method just described,
at the location of all isolated (no companions within a radius
of 15 pixels), high S/N and unsaturated stars - a list comprising 220
stars. Each kernel is 15x15 pixels in size, which is significantly 
larger than the typical PSF with FWHM in the range 2-4 pixels. Since
the PSF's to be matched have essentially the same functional form, the
largest value in this array is almost exclusively located in the 
central pixel.

To determine the behavior of the overall spatial variability of the 
kernels a polynomial 
surface fit was made to the value of this pixel at the $x,y$
positions of each kernel determination on the frame. Figure 3 shows a 
contour plot
of the result of a typical high order polynomial fit for a frame 
pair. In this example the fit excludes the smaller region 
marked by the inner contour with a value of unity where the relative 
size of the PSF on each frame reverses. The unit contour represents
the location where the PSFs are close to identical. In Figure 2 this
would be located where the FWHMs radial forms cross at a radial
distance of $\sim 450$ pixels.

\begin{figure}
\plotone{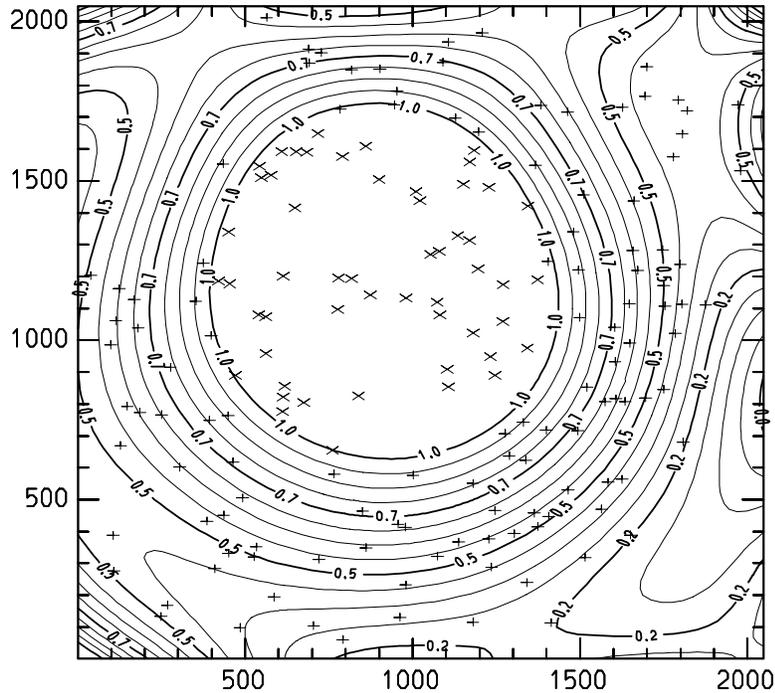}
\caption {
The spatial variability of the PSF matching convolution
kernel for a pair of images in the KPNO 4-m data.
The empirical PSF matching convolution kernel
has been measured at the location of points marked on the plot. The
$\times$ marks are distinguished from $+$ marks as locations where
the ratio of the size of the PSF on the two frames either exceeds or
is less than unity.
A 6th order polynomial surface fit (as a function of CCD coordinates)
has been made to the central pixel intensity
of the matching convolution kernels
at the $+$ mark locations and is shown as a contour plot. The inner
contour marks the region where the PSFs are close to the
same size on the original frames and within this region a similar
determination of the PSF matching convolution kernel is made but with
the images reversed.
\label{fig3}}
\end{figure}

Since the normal variation of the PSF within these frames generally 
exceeds the variation in seeing between frames the situation in 
shown in Figure 3, with an inner elliptical or near circular
region within which the relative size
of the PSF's of the frame pair reverses, is quite typical. The 
overall radial variability of the convolution kernel in both 
regions is always concentric about the same point, however the 
position of this point was found to vary up to 100 pixels between frames
from the symmetry point apparent in Figure 3. The location of this point 
most likely represents the principal axis of the telescope which is 
slightly displaced from the center of the CCD and telescope flexure
may account for the movement of this point from one frame to another

Spatially dependent kernel models were derived from
($r,\cos\theta$) polynomial fits to each pixel, $i,j$, in the 15x15 array
of the 220 raw kernels derived for each frame. 
The radial variation of the PSF is by far the dominant term, but 
models were generally better fit with
some azimuthal component, invariably with symmetry about an
axis $\theta_o$. Typical models were
parameterized by locating the radial point, ($x_o,y_o$),
and an azimuthal symmetry axis
from contour plots such as Figure 3.
The general fit is given by:

$$k_{i,j} (x(r,\cos\theta),y(r,\cos\theta)) = 
\sum_{n,m} {a_{i,j}(n,m) ~r^{n} \cos^{m}\theta}, \eqno(4) $$

\noindent
where $r$ is the radial distance from the PSF symmetry point and 
$\theta$ measured with respect to the azimuthal symmetry axis.
The best fits were cubic ($n=3$) in $r$ and linear ($m=1$) 
in $\cos\theta$. Prior to the overall fit an
interactive examination of the radial plots of the central kernel
pixel was performed in a small number of azimuth sections in order to 
delete any outlying raw kernels before the final fit. 
In the situation illustrated in Figures 2 and 3, with two regions 
defined by the ratio of the sizes of the PSF on each frame,
two separate model fits were made. 

\subsection{Image Subtraction}

Before subtraction both the image and reference image were processed
further. Images were ``unsharp masked'' - 
the underlying smoothed galaxy and sky background were removed
from both images by subtracting the large-scale median smoothed image 
of both frames. This leaves only the data to reference frame photometric 
scaling factor to be determined.
The measures of the 220 raw kernels were used to do this. First, 
a 10x10 pixel box around each of the PSF stars was matched, then the
intensity scaling factor was derived from a linear fit to the 
data versus reference pixel intensity match within this box
(using software provided by A. Phillips). 
The scaling factor for the entire data image was the
median value of the independent scaling factors determined for the
220 PSF stars, rendering it insensitive to any stellar variability,
and was accurate to 1.5\%.

As a convenient method of performing later photometry, a well spaced grid 
of 16 of the 220 PSF stars were removed from the data frame by zeroing the 
pixel values in a 16x16 pixel box centered on each of these stars. Since
this step is performed prior to overall PSF-matching of the data and
reference image and the subsequent subtraction, this
has the advantage of ensuring properly scaled, high S/N and 
PSF-matched reference stars from the reference image
on the final subtracted image for the
purposes of differential photometry.

For computational efficiency the models derived from equation (4)
were used to compute kernels for a grid of $(r,\cos\theta)$ positions.
In matching a frame to its reference image, the frame was divided up into an
$(x,y)$ grid of subimages and the nearest kernel to the center of 
each subimage from the model grid was used to perform the PSF-matching 
convolution for that subimage. The resolution of both the model kernel
grid in $r$ and $\cos\theta$ and the image processing $x,y$
grid were chosen to be 
at least at the point where no significant improvement 
in the subtraction quality could be obtained with increasing resolution.
For the kernel model this was typically around $\Delta r$ of 50 pixels
and $\Delta\cos\theta$ of 0.2.
The final image-reference differenced subimage comprises
either a data frame PSF-matched to its reference frame or a reference
frame PSF-matched to the data frame,
depending on the ratio of the sizes of the PSF's 
of the data and reference image within the subimage.
In {\S 4} we assess the quality of the subtraction we have achieved with
this algorithm.

\subsection{Simplifying the PSF Matching Convolution Kernel with an
Optical Corrector}

If an observer utilizing image differencing techniques outlined
above has influence over
the choice of telescope/imager optical design, a great deal of effort can be
saved in the data reduction stage by using optics that produce a uniform PSF
over the entire detector, and in particular an optical system that produces a
final focus coincident with the detector surface. 
Reducing the spatial dependence of the PSF
correspondingly reduces the effect of changes in telescope focus
inducing a spatial dependence of the full-frame PSF-matching kernel.
In the
case of our VATT survey, we installed a doublet biconvex achromat, designed
with the aid of Richard Buchroeder of Tucson, Arizona, that produces uniform
20 micron ($0.25''$) 
diameter spots and best focus over the entire surface corresponding
to the curved SITe 2048 backside-illuminated CCD (with a curved surface of
approximately 220 microns sagitta center-to-edge). This resulted in 
$1-4$ PSF kernels
being required for the VATT image, versus approximately 200, 
together with the modeling algorithm in {\S 3.4}, for
the KPNO 4m prime focus
CCD data, which covers only 2.1 times the solid angle 
in field of view.

\section{TECHNIQUE SENSITIVITIES AND SYSTEMATIC EFFECTS}

In this section we assess the sensitivity of the technique and 
discuss systematic effects that are generally pertinent to DIP
and the surveys we have outlined.

\subsection{Difference Image Photometric Errors for Faint Unresolved Stars}

The application of the algorithm outlined in {\S 3} to the KPNO data
was successful at yielding good subtraction over most of
the frame, even in the unsaturated regions of the 
inner part of the bulge which had proved very difficult with other methods.
In Figure 1 we show an inset subimage of the central region of
the poorly differenced subimage,
but after the full-frame PSF matching function model has been
applied. This illustrates the importance of accurate PSF matching: a
clear detection of a point source can now be seen in the frame, which was
previously almost completely undiscernible
in the systematic residuals in the original
attempt at matching the PSF ({\S 3.2}).

The PSF-matching starts to break down at large values of $r$ where the PSF
is varying most rapidly. The radial symmetry point of the PSF matching
function was located
close to the center but always in the same quadrant (the top left quadrant
as seen in Figure 3). 
Consequently, the poorest subtraction is always at the furthest radial
distance in the diametrically
opposite quadrant of the frame ($r > 1100$ in Figure 2 which 
corresponds to the bottom right corner in Figure 3)
where an insufficient number of 
PSF stars can be found to constrain the model fit. However,
the image subtraction is quite satisfactory over most of the CCD 
excluding saturated regions, which we discuss below.

The effectiveness of the subtraction can be quantified by measuring the
ratio of the standard deviation of the residuals in the subtracted frame
to the predicted noise based on the photon and read-noise for the region
of the frame being examined. This can be done simply by constructing
a ``noise'' image for a particular difference frame to be examined. 
Since the read-noise of the CCD is 
negligible in all regions of the well exposed frames the predicted
noise is determined only by the photon noise. Such a noise image, $n$,
is given by the following addition of the image $i$ and its reference
image $r$,

$$ n = ({1\over g} (s_i^2 {i\over N_i} + {r\over N_r}))^{0.5}, \eqno (5)$$

\noindent
where $N_i$ and $N_r$ are the number of images comprising $i$ and $r$
respectively if these images are averages of frames, $s_i$ is the 
photometric scaling factor required intensity match $i$ to $r$, and
$g$ is the fixed gain of the CCD in electrons per ADU and can be
measured from flat-field images at the telescope. Dividing a difference
frame by image $n$ generates a difference image where the residuals
are given in units of the predicted photon noise in each pixel.
Since one of the two images has been degraded by convolving with the
matching PSF kernel, at a given location 
the photon noise per pixel is reduced by the 
averaging effect of the kernel. Thus equation (5) can overestimate
the photon noise, however, in a typical situation where ($N_r >> N_i$)
and the seeing is better in the reference frame, the noise is effectively
completely determined by the photon noise in the $i$ frame which has not
been convolved.

Histograms of the residuals of typical difference 
frames in units of the photon noise, $\sigma_{photon}$, show
residuals of $< \pm 8\sigma_{photon}$ over 90\% of the frame.
However, about 5\% of the original image is 
saturated (mostly in the nucleus) 
and the smoothing of the PSF-matching convolution 
kernel leads to $\sim 2\%$ of contamination in the difference frame around
saturated stars (this effect can be seen in Figure 1). 
Over most regions where there are no 
obvious systematic residuals due to bright stars the standard deviation
in the difference frame is typically within $1.5\sigma_{photon}$.
The pixel-to-pixel residuals do not quantify the quality of the 
subtraction, however, since systematic errors in
the subtraction technique lead to residuals 
that are correlated on the scale of the PSF. 
Convolving a difference image with a boxcar of size N$\times$N and 
dividing by a noise image scaled by N is a convenient way to examine
noise on larger scales. Figure 4 shows how the residuals do indeed
become larger in terms of the photon noise on bigger scales. For N=1,
corresponding to the pixel-to-pixel scale the standard deviation is
$1.8\sigma_{photon}$, but for N=5,
which corresponds closely to the size of the typical PSF in these
data, the difference frame residuals are $2.4\sigma_{photon}$ and 
increase to $4.4\sigma_{photon}$ for N=20, in this example. 
For large N this increase over the
entire frame appears to be due to systematic effects such as fringing
which is a problem in I-band CCD photometry. An examination of the 
residuals in smaller ``clean'' 
subimages shows that the standard deviation does
appear to return close to $\sigma_{photon}$. However, the only scale 
we are concerned with is the PSF size, and here we find almost all of
our difference frames typically achieve residuals in clean, unsaturated
regions that are fully consistent standard deviations of 
$2\sigma_{photon}$ and range up to $3\sigma_{photon}$.

\begin{figure}
\plotone{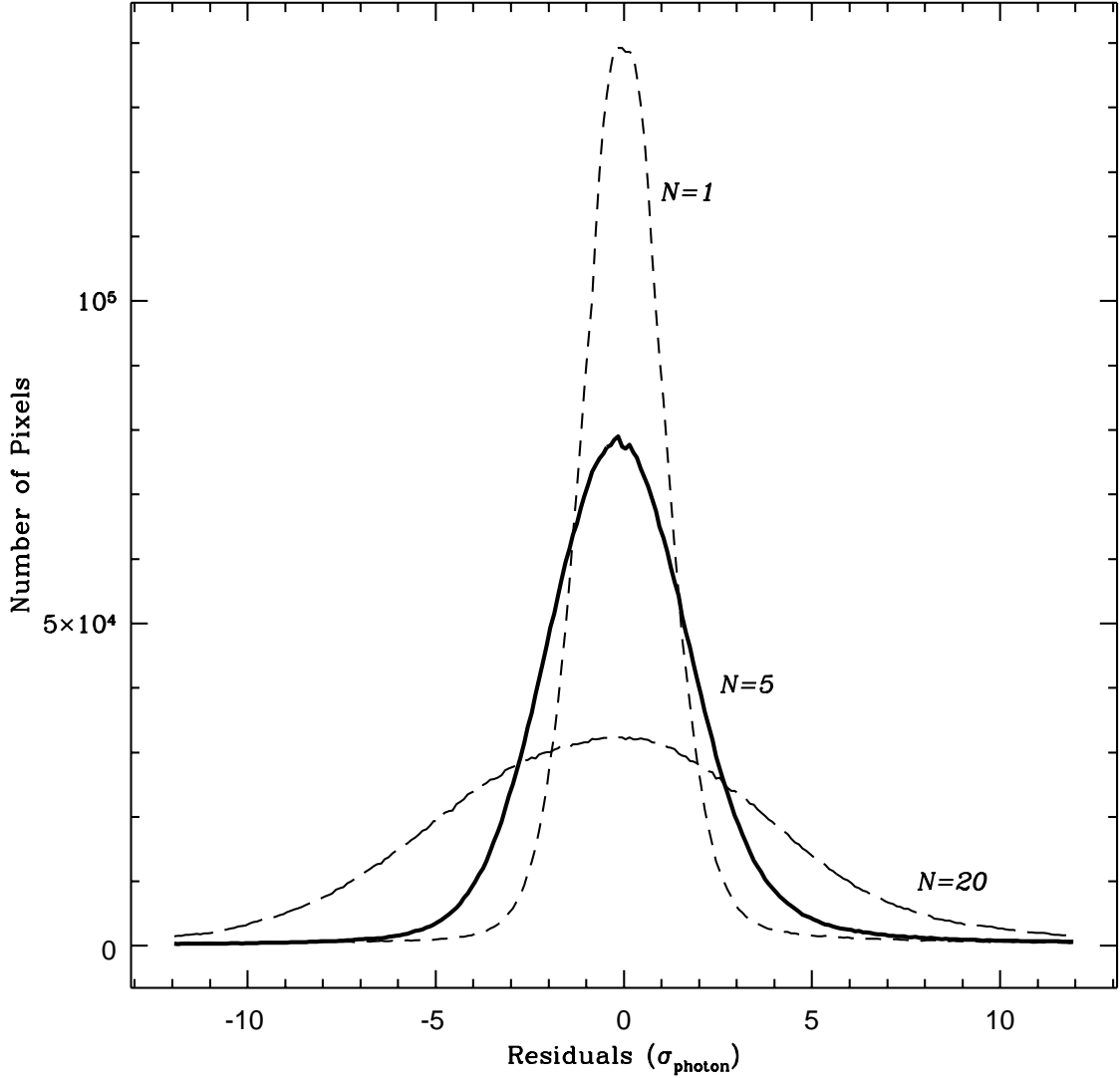}
\caption {
A histogram of the residuals in a typical full-frame difference
image expressed in terms of the photon noise and calculated for
differing scales of pixel smoothing. For the pixel to pixel (N=1) scale
the standard deviation of the curve is $1.8\sigma_{photon}$. However,
due to various systematics ({\S 4}) residuals are correlated on the scale
of the PSF (N=5) which is reflected in its curve which shows a standard
deviation of
$2.4\sigma_{photon}$. On larger scales for the full image
the standard deviation
continues to increase, but this is mostly due to large
scale fringing in this case of this I band difference frame; examination
of smaller subregions of these images actually show residuals consistent
with $\sigma_{photon}$. Most importantly, however,
almost the entire frame has residuals within a factor of three of
the photon noise on the scale of the PSF ({\S 4.1}).
\label{fig4}}
\end{figure}

Expressing the residuals in the difference frame in terms of photon
noise quantifies
precisely how the differencing technique presented here compares
with the ultimate noise limits. 
Once the true photometric errors are 
determined by measuring the standard 
deviation of residual pixels binned on the scale of the PSF it 
appears DIP is quite
competitive with conventional photometry of faint resolved stars.

\subsection{Difference Image Residuals of Bright Stars}
 
It is possible that some ``sources'' in difference frame are simply
due to residuals of bright, resolved stars. To verify
that candidates were not systematic residuals of bright stars, photometry
was made at the location of candidates in the unsubtracted frames.
The validity of the subtraction
for a particular candidate is easily confirmed by comparing the residuals
of a neighboring bright star in the noise regime $\Delta F_{err} >>
\sigma_{photon}$, where $\Delta F_{err}$ represents the systematic
error in the
subtraction of a star of flux $F$ and both terms are calculated over the PSF.
In this noise regime the fractional error $\Delta F_{err}/F$
for individual bright
stars may vary over the frame depending on the quality of the subtraction.
For candidates with stars of flux $f < F$, but bright enough to
be resolved in the raw frame, we can
use a local estimate of the fractional error in the subtracted flux to
distinguish between artifacts of the subtraction
and real variability, where
$\Delta f/f  >> \Delta F_{err}/F$.
 
\subsection{CCD Photometric Accuracy and Surface Brightness Fluctuations}

In {\S 3.5} we determined that the photometric scaling 
factors determined from our PSF matching starlist was
accurate to 1.5\%. If the photometric accuracy of these
stars was determined entirely by photon noise these scaling
factors should be accurate to 0.1\%, indicating that
systematic errors must account for their empirical
accuracy. Flat-fielding may account for most of this discrepancy.
On the pixel-to-pixel scale care was taken to ensure that nightly flats
were of sufficient S/N that flat-fielded object frames and subsequent
combined frames contained little contribution of photon noise from
the flats themselves, but on larger ``illumination'' scales
we detect variability between domeflats around 1\% over hourly
timescales, which may be due to a small time variability in the 
linearity of the CCD. However, our empirical accuracy, due to this
or other reasons, is quite typical for CCD work and here we assess how
this intrinsic accuracy influences the residuals in
our difference frames. 

The structure seen in the unsubtracted subimages in Figure 1 and 
similar exposures of nearby galaxies is due to the statistical 
fluctuation of the number of stars in the galaxy contributing to
each pixel or seeing element (Tonry \& Schneider 1988). The
amplitude of these surface brightness fluctuations (SBFs) is
determined by the luminosity function (LF) of the stellar population
and the surface brightness and distance of the galaxy. For a given
seeing element this amplitude, $m_{SBF}$, can be expressed in terms 
of the empirical measure of the ratio of the second and first 
luminosity moments of the LF, $\bar m$ (the fluctuation magnitude). 
Following the formalism of Gould (1996) this is given by,
 
$$ m_{SBF} = (-2.5 {\log [(S-S_{sky}) \Omega_{psf}]} + \bar m
+ m_{1})/2, \eqno (6) $$
 
\noindent
where $S$ and $S_{sky}$ are the surface brightness in counts per 
pixel of the galaxy plus sky and sky respectively, $\Omega_{psf}$ 
is the size of the resolution element in pixels which we adopt to be
$\pi {\rm (FWHM)}^2$, and $m_{1}$ is the magnitude corresponding to
1 ADU on the CCD frame.
 
The observed $\bar m_I$ for M31 is 23.29 (Tonry 1991 and using his
adopted foreground Galactic extinction of $A_B = 0.31$). The 
(V-I) color of M31 is 1.18 (Tonry \etal 1990), and using galaxies
of comparable color from this work, we estimate 
the observed $\bar m_R$ to be 23.99. 
For the mean seeing of $1.5''$ for our two best nights and the
calibration of the surface photometry of our frames ({\S 5}),
the mean value for $m_{SBF}$ is 20.8 for the R frames. The 
1.5\% photometric scaling accuracy of this mean SBF predicts an
error of R=25.5 in the difference frame which is almost identical
to the photon noise in the same size seeing element for the
mean surface brightness for the R frames ({\S 2.2}). Thus the intrinsic
accuracy of the photometric scaling accounts for $1.4\sigma_{photon}$ 
seeing element residuals in the difference frame.

\subsection{Effects of Atmospheric Dispersion in a Difference Frame}

The effects of atmospheric refraction can become
a serious issue when applying DIP.
Filippenko (1982) tabulates atmospheric dispersion for an
observatory at an altitude of 2 km (comparable to KPNO)
as function of airmass in the wavelength range
3000-10000{\AA}. Our main concern is the effect of the atmosphere on
the centroid of a star as a function of airmass in a broad bandpass filter.
Photons at the wavelength extremes of our R filter (5600 to 7200{\AA}) 
separate out from $0.00''$, at an airmass of 1.0, up to $0.65''$,
at an airmass of 2.0 within the stellar image (for the extremes of the 
I band filter in {\S 2.1} the separation extends to $0.40''$ at 2.0
airmasses). 
This has the effect of elongating the stellar PSF
in the direction of atmospheric dispersion at higher airmasses, leading
to poorer image subtraction when differencing frames with different
effective airmasses. If all the stars on the frame are the same color 
then this elongation affects the shape of the PSFs in the same 
way. The Fourier algorithm ({\S 3.2}) can still cope with these
effects on the PSF when matching frames taken at different effective
airmasses and automatically correct this problem. 
However, the individual colors of stars in a frame lead to
second order effects: the centroid and elongation of a star's PSF
will differ depending on its color. There is no easy way to compensate for
this, particularly without knowledge of the precise position and color of
all detected stars on the field.
Such an effect will yield poorer subtraction with increasing airmass and
result in power on the scale of the PSF in the difference frame.

Fortunately, M31 is at a favorable declination for mid-latitude sites 
in the Northern hemisphere
and can be accessed for roughly eight hours a night at airmasses
$< 1.5$, and this problem is quite small for our selected bandpasses.
Furthermore the RG stars we are sensitive to ({\S 2.1}) have red 
colors of $\langle (V-I) \rangle = +1.2$ (the mean color in the bulge region,
Tonry 1991), which makes the 
PSF distortion smaller compared with bluer stars.
All the KPNO data presented here were taken at $< 1.5$ airmasses and
no discernible degradation was seen in the difference frame residuals 
of frames taken at higher effective airmass.
However, we note that the effect can be quite large for shorter
wavelength bandpasses. 
In the wavelength extremes of a B filter: 3500 to 5000{\AA}, for example, 
the centroids separate by $2.''1$ between 1.0 and 2.0 airmasses.

\subsection{Signal Contribution of RR Lyrae Variables}

RR Lyrae stars will be in abundance in the region of the bulge we
are surveying and these stars 
have been observed by Pritchet \& van den Bergh (1987) in the halo 
of M31 40$'$ (9kpc) from the nucleus.
Such variables have periods in the range $1.5-24^h$ and thus it
is important to consider their possible contribution to the KPNO data which
is sensitive to these timescales. Pritchet and van den Bergh 
determined a $\langle B \rangle$ of 25.68 for 30 RR Lyrae candidates.
Adopting a mean
$\langle B-V \rangle$ of 0.26, which corresponds roughly to a G0 star (Hawley 
\etal 1986) and a Galactic foreground extinction of $A_B = 0.31$ 
(Burstein \& Heiles 1984), their $\langle R \rangle$ should be $\sim25.2$.
The maximum observed amplitude range of $1.6^m$ in $V$ 
(Bono \etal 1995) and so an RR Lyrae
star may be seen in a difference frame as a source no brighter than
R = 24.7. Given the construction of our KPNO survey ({\S 2.2}) 
this would correspond to a maximum S/N of $2.1\sigma_{photon}$ in the 
difference frame (and fainter in the I band difference frame). 

Thus RR Lyraes are just at the limits of our 
sensitivities in the KPNO survey given our assessment of DIP
in {\S 4.1}. However, it is quite possible that the 
correlation of difference frame residuals on the scale of the 
PSF discussed in {\S 4.1} contains some contribution from these stars in 
these surveys. It is therefore important to estimate
the RR Lyrae specific incidence, since RR Lyraes can be a significant
source of random noise in the photometry. Pritchet \& van den 
Bergh (1987) estimate for the halo of M31, 
the specific incidence (per integrated flux of the entire 
population) is about 100 Lyraes for each B=15.9 (R $\sim 14.5$,
from Walterbos \& Kennicutt 1987). In comparison, the total 
integrated magnitude over our KPNO field is R = 4.67, corresponding
to 850,000 RR Lyraes (or 0.9 RR Lyraes per square arcsec, on average),
{\it if} the specific incidence of RR Lyraes is the same in the halo
and bulge/inner disk.
If instead, the bulge/inner disk metallicity is much
higher than that found for the halo,
the bulge/inner disk could have up to 100 times fewer RR Lyraes than
predicted here. However, RR Lyraes might be a 
significant source of fluctuations on several hour timescales, which
is a supposition we will test in a later paper by studying the 
temporal frequency of variations of marginally detected sources in the 
difference frame.

\subsection{General Comments on PSF Matching} 

Critical to the techniques presented in {\S 3} is the assumption that 
the PSF is sampled at least at the Nyquist frequency (roughly $> 2$ pixels 
FWHM). This was the situation in all the data presented here. If the PSF
is not Nyquist sampled then this leads to irretrievable loss of information
concerning the PSF. In particular, registration of frames to common 
coordinates will lead to systematic ``resampling'' 
errors on the pixel scale that are
particularly obvious around bright stars ({\it c.f.,} Gould 1996). Such 
systematic effects can be reduced by averaging registered, undersampled
frames at the price of time resolution. 
Also, poor sampling similarly affects the Fourier determination
of the matching convolution kernel. By averaging empirically determined
convolution kernels of many stars the systematic errors in the convolution
kernel determination can be reduced, but given possible limitations in
the number of PSF matching stars on a given frame ({\it c.f.,} {\S 3.3})
this may not be a viable option. As with registration of undersampled data,
even with an accurate matching convolution kernel the undersampling will
still lead to systematic effects in the PSF matching, which would also 
best be coped with by averaging multiple difference frames.

The Phillips \& Davis (1995) 
algorithm ({\S 3.2}) makes an assumption about the
behavior of the wings of the PSF in the replacement of high frequency 
noise dominated components with a fit based on an elliptical Gaussian. 
The difference between the ideal kernel and the empirical kernel 
determined by this method will reflect the difference between the real 
behavior of the wings of the PSF and the assumed model. These differences
will propagate as systematic errors in the difference frame that are
correlated on the scale of the PSF. If the behavior of the wings of the
PSF is well known, then in principle the algorithm can
be modified by fitting the wings of the PSF with a more realistic model.

Even with a relatively uniform PSF across an image a spatial dependence
of the PSF matching convolution kernel can be induced
when trying to match images
that are not taken at close to the same position in the sky. This is 
because of astrometric distortion in the image plane of the detector. 
Frequently in a wide field imager
the transformation between frame coordinates is not a simple linear
shift, rotation and magnification; it contains higher order terms. The
IRAF routines, GEOMAP and GEOTRAN, allow for such higher order terms
to enable accurate registration, however, the non-linear transformation
will result in a {\it different} 
spatial dependence of the PSF in the registered frame compared
to the unshifted image it is to be differenced with.
Consequently, this
has the effect seen in the KPNO data ({\S 3.3}) - a spatially 
dependent PSF matching convolution kernel is required to match frames 
and can be treated with the algorithm discussed in {\S 3.4}. 
(In test frames we took at the McDonald Observatory 2.7-m in 1994, we
found precisely this effect with
significant degradation in the quality of differenced images 
comprising an image pair that originally differed in registration
by 15\% of the size of the CCD, despite the uniform PSF
across the $3.6'\times 3.6'$ field of view of the camera.)
This problem and its solution is also relevant to 
differencing images taken on different telescopes.

In sum, the factors
discussed in {\S 4.3} to {\S 4.6} all contribute to the correlated 
residuals on the scale of the PSF in the difference frame and some of
these have been assessed analytically by Gould (1996). 
Once a candidate is determined not to be a residual of a bright star in 
the difference frame ({\S 4.2}), then given all of these
systematic effects, the reality of a detection
of variability is best determined by measuring 
the variance of the 
residuals {\it integrated on the scale of the PSF} around the candidate
source in the difference frame, since this measures the true photometric
errors in DIP. The errors in the light curves of all objects
presented in {\S 5} have been determined this way.

\section{RESULTS}

The analysis was on performed on R and I sequences
comprising eight and nine KPNO 4-m prime focus 
images which were themselves
combinations of five consecutive images each for the nights of September
26 and 27. Since conditions compromised the quality of the data on the
previous two nights ({\S2.2}), all the data on these nights were combined
into single averaged frames of comparable S/N to the individual combined
frames on the last two nights. Thus the total number of averaged
frames analyzed for each filter was 19.

Photometric calibration of both the R and I filters was made with 
images of M92 taken under photometric conditions on the last night of 
the run. Images taken that same night at the same airmass as the M92
observations were used to make the final calibration of the M31 fields.
We used the photometry of Christian \etal (1985) in their 
R and I filters. Our broad band R filter has a bandpass 70\% wider
than a Kron-Cousins CCD R filter (Schild 1983), but essentially the
same effective wavelength ($\lambda6470$). The I filter we use 
has an effective bandpass of 3565{\AA}, which is about twice the
Kron-Cousins equivalent, and an effective wavelength of $\lambda8800$,
compared with $\lambda7990$ for the Kron-Cousins I. In both filters
the scatter in our raw magnitudes and the Christian \etal R and I 
equivalents was better than $\pm 0.^m1$ for 10 stars with (V-I) colors
in the range $0.^m0$ to $1.^m4$. The agreement with a Kron-Cousins I 
filter, despite the differences in the photometric bandpass with our
I filter, probably reflects the falling CCD sensitivity at redder
wavelengths that will tend to match the overall detector response 
to the Kron-Cousins I. 
We have not attempted to correct magnitude
estimates with color terms in our photometric transformations so 
our photometric calibration is only accurate to $\pm 0.^m1$.

As we discuss below, it is useful to compare sources in the raw and
difference frames to the local SBF amplitudes. The amplitude of the 
SBF in a seeing element is given by equation (6), which requires 
the surface brightness of the galaxy above the sky to be determined.
We have used an estimate of the sky brightness in our photometric
moonless R frames of R = 20.8 per square arcsecond. (This is only 
$\sim 50\%$ of the total surface brightness 20$'$ from the nucleus
- the galactocentric extreme of our field.) This sky estimate gives
surface brightness values consistent with the photometry of Walterbos
\& Kennicutt (1987) in R. 
In the case of the I filter variable night sky OH line
emission makes the sky estimate much more uncertain. Since M31 does
not exhibit a strong color gradient in this galactocentric range
we solved for the I sky
brightness for one of our I frames by determining the sky value that
gave the minimum scatter in the (R-I) galaxy color 
for locations scattered throughout the field. A sky brightness value
of I $\sim 19.2$ for this typical photometric, moonless frame gave a mean
(R-I) $= 0.80 \pm 0.03$ for the entire field and I surface brightness
values consistent with the photometry of Hiromoto \etal (1983). For our
purposes the accuracy of these estimates is quite sufficient. 

All the KPNO frames for the two bandpasses 
were differenced against their respective reference frames 
of the combined images from the final night ({\S 3.1}) after the 
processing outlined in {\S 3}. The search of each frame for sources
was conducted by visual inspection of the difference frames. 
Many sources were initially detected in the difference frame of the 
combined images for the two last and best nights of the run. Thus the 
search is very sensitive to changes on the night to night timescale. 

We used IRAF DAOPHOT to perform photometry of the sources on the difference
frames. To minimize the problems of the variable PSF photometry of a source
was made relative to neighboring calibrated bright stars from the PSF
matched reference frame. 
If the source in the difference frame has negative flux we applied DAOPHOT
to a ``negative'' difference image.

To highlight the sensitivity of the image differencing technique 
Figure 5 shows two consecutive combined $36''\times 36''$ R band
subimages separated by $50^m$ in time taken on the last night. The
upper panels are the undifferenced images, but with the large scale 
median smoothed galaxy and sky background subtracted. The lower 
panels are the result of differencing the original images with a
reference image comprising an average of all images taken on the 
previous night. The difference image for the second frame shows 
a clear detection ($20\sigma$) of variability over the previous frame.
Remarkably, the eye cannot discern any indication of such variability
in the raw frames. The reason for this can be clearly understood 
in terms of the underlying surface brightness fluctuations associated
with the unresolved stars in the galaxy. The differential flux
in the difference
frame has a magnitude of R = 21.14. Correcting for an estimated sky 
with a surface brightness of R = 20.8 per square arcsec, 
we measure the galactic
surface brightness to be R = 18.29 per square arcsec
at this location (0.5 kpc along the far side minor axis). With $1.2''$ 
seeing and applying equation (6) the
predicted surface brightness fluctuations in a seeing element have
an amplitude of R = 20.32. Thus the SBFs at this location 
have an amplitude that is 
more than twice as bright as the total flux of the detected source in the 
difference frame. 
 
\begin{figure}
\plotfiddle{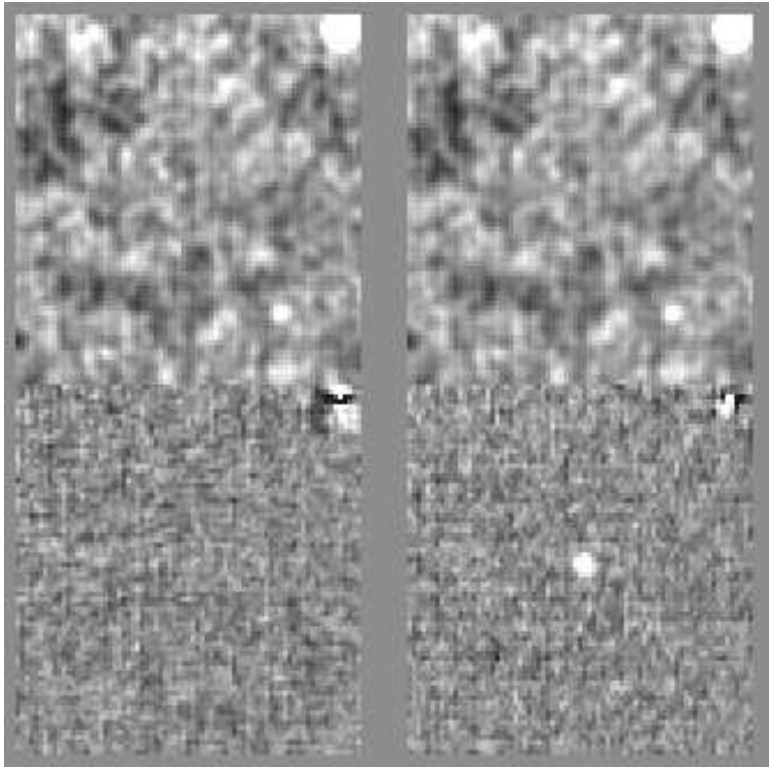}{4in}{0.0}{50}{50}{-288}{-648}
\caption {
Image differencing at work:
the sensitivity is dramatically illustrated
in this image in which shows the raw (upper) and difference (lower)
$36\times 36''$ subimages of
two consecutive R-band images taken 50$^m$ apart. The raw images
have been stretched to maximize as much as possible the surface
brightness fluctuation structure in the unsubtracted data. The difference
images are stretched at 10\% of the upper panels to highlight the
noise in the difference frames and have been created by subtracting
a high S/N reference image from the previous night after suitable
registering, scaling and PSF-matching.
The bottom right panel shows a clear $20\sigma$ detection of
variability relative to its previous image. Despite this large
amplitude it is still completely invisible to the eye
in the top right unsubtracted image. This is due to the fact that
in the $1.2''$ seeing it is only a 1.5\% modulation of the background
galaxy light which has a seeing element flux of R = 16.3.
This amplitude is still only half the amplitude of the surface
brightness fluctuations at its location ({\S 5}). It is the ability of
image differencing to be sensitive to flux variations in arbitrarily
high stellar crowding that expands enormously the number of stars
a microlensing survey can be sensitive to once such a survey
is no longer limited by whether the stars are actually resolved. The
light curve of this object is illustrated in Figure 7.
\label{fig5}}
\end{figure}

The flux difference light curves in both R and I of this source 
(candidate \#1 in Table 1) can be 
seen in Figure 6. The subimages in Figure 5 correspond to
the 14th and 15th data points of the R band curve. It is likely 
that this particular variable is a nova. 
The nova rate in M31 is well determined: 
26 novae year$^{-1}$ (Arp 1956; Capaccioli \etal 1989) and the spatial
distribution closely follows the bulge light (Ciardullo \etal 1987). 
Thus there is probability of $\sim 10\%$ that one will erupt during our
four night run, since our field covers roughly one quarter of the 
bulge light. Our final measurement of this object $3.5^h$ after its
initial detection in the difference frame has a difference magnitude
of 19.33 relative to the previous night. If this is a nova it may
have peaked at an R magnitude brighter than $17.^m0$. 
A number of novae are undoubtedly present in these frames since novae of all
speed classes intersect in brightness at 15-18 days past maximum light
(Buscombe \& deVaucouleurs 1955)
with an absolute magnitude of $M_V = -5.6$ (Capaccioli \etal 
1989) which corresponds
to R $< 18.5$, well above our detection thresholds.

\begin{figure}
\plotone{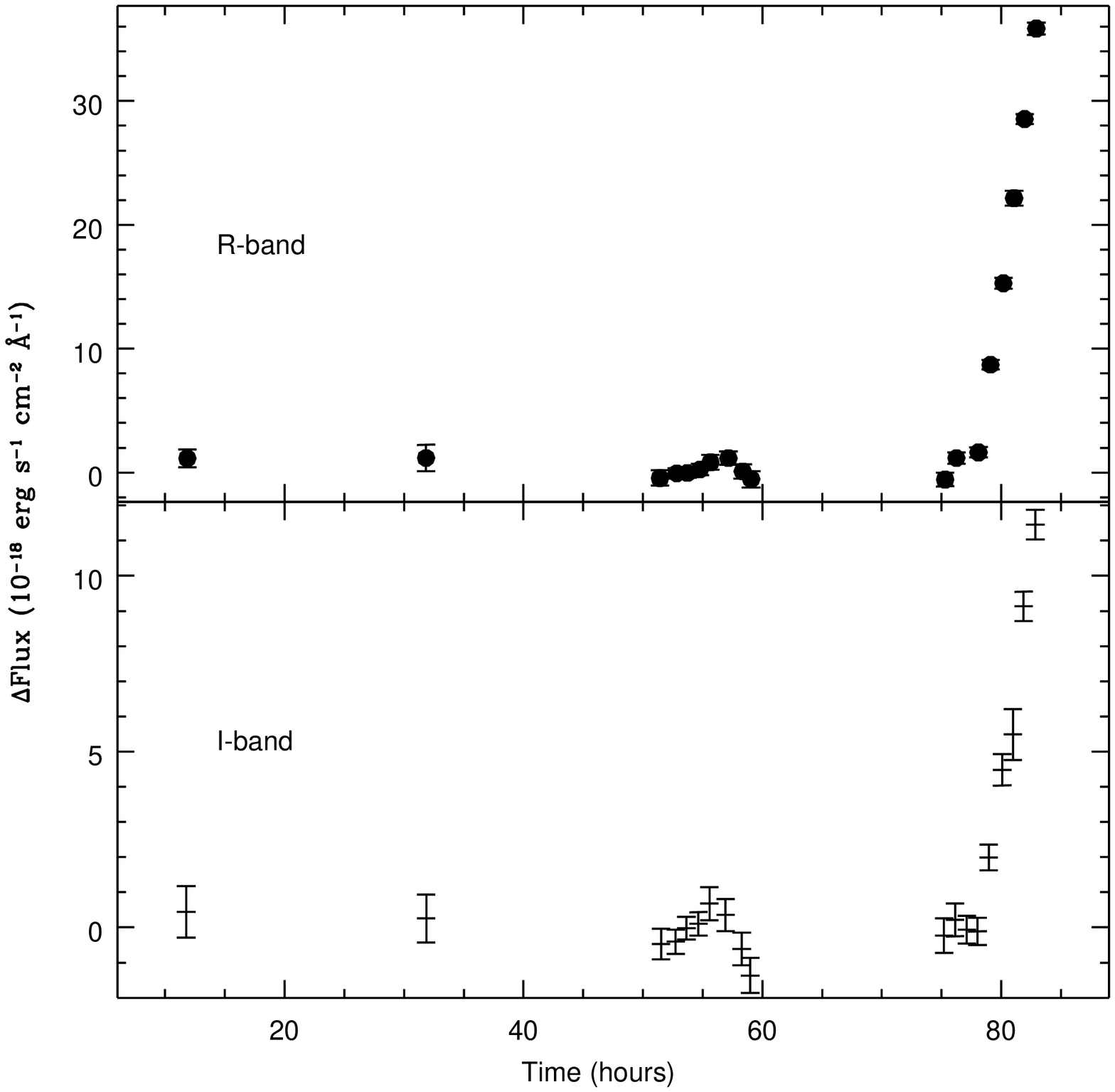}
\caption {
The light curve of the rapidly varying object in Figure 5
The differential fluxes are all measured relative to
the mean image of the third night in both filters. Points 14 and 15
of the R light curve correspond to the subimages in Figure 5. This
object is most likely a nova which is just caught in eruption.
The reality of the apparent variability during the third night is
not clear, since this object was very close to the heavily
saturated nuclear region of
the CCD where some difference images exhibited a ``plaid'' structure
(not seen in Figure 5), apparently an occasional
artifact of the CCD read-out at high signal.
The Julian Date at $t$ = 0 is 2,449,619.50.
\label{fig6}}
\end{figure}

\makeatletter
\def\jnl@aj{AJ}
\ifx\revtex@jnl\jnl@aj\let\tablebreak=\nl\fi
\makeatother

\begin{deluxetable}{lllrrlllll}
\tablewidth{43pc}
\tablecaption{Detected Sources}
\tablehead{
\colhead{ID}                & \colhead{RA (2000)}         &
\colhead{Dec (2000)}        & \colhead{R}                 &
\colhead{I}                 & \colhead{$\Delta$R$_{max}$} &
\colhead{$\Delta$I$_{max}$} & \colhead{$\mu_R$}           &
\colhead{$\mu_I$}           & \colhead{Remarks}
}
\startdata
001&0 42 58.6&41 15 50&$>$19.7&$>$18.9&19.58&19.68&18.29&17.48&nova? \nl
002&0 42 53.9&41 04 46&20.64&$>$20.0&21.31&21.29&20.54&19.75&eclipsing? \nl
003&0 42 58.9&41 12 10&$>$20.3&$>$19.5&21.91&21.69&19.46&18.69&~ \nl
004&0 42 57.4&41 11 37&$>$20.3&$>$19.6&22.89&22.06&19.55&18.77&~ \nl
005&0 42 58.3&41 11 20&20.02&18.97&21.67&21.38&19.67&18.92&~ \nl
006&0 42 58.8&41 11 03&19.72&18.74&22.28&22.37&19.80&19.02&~ \nl
007&0 43 11.2&41 10 31&$>$20.6&$>$19.9&22.23&22.15&20.18&19.40&~ \nl
008&0 43 14.5&41 12 31&$>$20.5&19.29&22.08&21.89&19.95&19.13&~ \nl
009&0 42 59.4&41 13 59&19.20&18.32&21.60&21.84&18.95&18.17&~ \nl
010&0 43 07.5&41 10 51&$>$20.5&$>$19.8&22.15&21.78&20.05&19.27&~ \nl
011&0 43 03.0&41 13 39&19.55&19.23&20.67&20.43&19.25&18.46&~ \nl
012&0 43 23.0&41 10 55&20.33&$>$20.0&22.24&22.16&20.37&19.59&~ \nl
013&0 43 24.3&41 10 50&$>$20.7&$>$20.0&22.20&22.03&20.44&19.63&eclipsing? \nl
014&0 43 25.0&41 10 32&20.03&19.93&22.09&22.10&20.44&19.67&eclipsing? \nl
015&0 43 30.2&41 10 36&$>$20.8&$>$20.0&21.90&21.92&20.53&19.74&~ \nl
016&0 43 29.6&41 14 12&$>$20.6&$>$19.9&21.19&21.31&20.18&19.40&~ \nl
018&0 43 28.0&41 13 55&$>$20.6&$>$19.8&21.84&21.90&20.16&19.34&~ \nl
019&0 43 28.7&41 13 43&$>$20.6&$>$19.9&22.10&21.93&20.24&19.41&~ \nl
020&0 43 26.2&41 12 02&20.43&19.38&22.27&21.93&20.35&19.56&~ \nl
021&0 43 27.6&41 11 14&$>$20.7&19.61&21.43&21.25&20.40&19.64&eclipsing? \nl
022&0 43 31.2&41 12 16&$>$20.7&$>$19.9&22.23&21.95&20.36&19.53&~ \nl
023&0 43 33.6&41 11 54&20.66&$>$20.0&21.58&21.44&20.49&19.73&~ \nl
024&0 43 43.6&41 11 49&20.42&19.74&21.51&21.24&20.79&20.02&~ \nl
025&0 43 44.9&41 11 49&$>$20.9&$>$20.1&21.96&22.26&20.71&19.95&eclipsing? \nl
026&0 43 45.8&41 11 51&20.20&20.08&21.86&21.66&20.76&19.96&eclipsing? \nl
027&0 43 49.0&41 11 32&$>$21.0&$>$20.2&21.42&21.46&20.87&20.07&eclipsing? \nl
028&0 43 50.7&41 12 25&20.78&$>$20.2&21.94&21.57&20.80&20.01&eclipsing? \nl
029&0 43 45.3&41 10 44&$>$21.0&20.18&21.97&21.62&20.88&20.04&~ \nl
030&0 43 37.3&41 14 17&19.99&18.95&20.79&20.46&20.41&19.56&~ \nl
031&0 43 37.6&41 12 05&$>$20.8&$>$20.1&21.80&21.42&20.61&19.77&~ \nl
032&0 43 42.9&41 10 57&$>$20.9&20.09&22.42&21.70&20.75&19.97&~ \nl
033&0 43 44.0&41 11 06&19.54&18.52&21.30&20.86&20.70&19.92&~ \nl
034&0 43 43.3&41 12 02&$>$20.9&19.58&22.47&21.82&20.76&19.95&~ \nl
035&0 43 48.2&41 12 56&20.16&19.46&21.19&21.46&20.67&19.88&eclipsing? \nl
036&0 43 56.4&41 13 04&19.59&19.00&20.67&20.52&20.82&20.04&~ \nl
037&0 43 50.4&41 14 18&$>$20.8&$>$20.1&21.38&21.38&20.66&19.80&~ \nl
038&0 43 52.2&41 14 14&20.42&$>$20.1&21.87&21.74&20.68&19.86&~ \nl
039&0 43 49.9&41 13 22&20.49&19.71&22.33&21.62&20.75&19.95&~ \nl
040&0 42 57.6&41 17 30&$>$19.6&$>$18.8&21.02&21.21&18.07&17.28&~ \nl
041&0 43 35.1&41 15 33&20.37&19.35&21.70&21.54&20.23&19.39&~ \nl
042&0 43 41.9&41 16 36&$>$20.7&$>$19.9&21.38&22.00&20.37&19.55&~ \nl
043&0 43 43.8&41 16 42&$>$20.7&$>$19.9&22.31&21.76&20.34&19.50&~ \nl
044&0 43 38.7&41 15 54&20.10&19.87&22.28&21.88&20.30&19.44&~ \nl
045&0 43 40.6&41 15 31&19.82&19.16&22.30&21.48&20.38&19.57&~ \nl
046&0 43 42.1&41 14 56&19.64&18.75&21.88&21.09&20.46&19.64&~ \nl
047&0 43 43.7&41 14 44&$>$20.8&$>$20.0&22.10&22.19&20.48&19.68&~ \nl
048&0 43 46.0&41 14 38&20.72&19.10&21.93&21.53&20.47&19.69&eclipsing? \nl
049&0 43 49.1&41 15 27&$>$20.8&$>$20.0&21.90&21.86&20.53&19.72&~ \nl
050&0 43 49.9&41 17 53&$>$20.7&$>$20.0&21.84&21.19&20.45&19.64&~ \nl
051&0 43 37.3&41 15 23&19.50&18.68&21.32&21.68&20.36&19.49&~ \nl
052&0 43 47.9&41 15 57&20.16&19.18&20.38&19.99&20.51&19.70&~ \nl
053&0 43 38.6&41 16 36&$>$20.6&19.65&21.66&21.78&20.22&19.39&eclipsing? \nl
054&0 42 55.7&41 05 45&$>$20.8&$>$20.0&22.46&22.01&20.51&19.70&~ \nl
055&0 42 58.9&41 05 10&20.72&19.46&22.14&21.21&20.57&19.79&~ \nl
056&0 43 06.8&41 04 35&$>$20.9&$>$20.2&22.77&22.35&20.78&20.00&~ \nl
057&0 43 05.0&41 03 58&19.66&19.02&22.88&21.58&20.76&19.94&~ \nl
058&0 42 56.9&41 03 13&20.55&$>$20.1&20.61&20.25&20.68&19.93&~ \nl
059&0 43 00.6&41 03 37&$>$20.9&$>$20.1&22.71&22.56&20.68&19.91&~ \nl
060&0 43 02.4&41 03 31&$>$20.9&$>$20.2&21.57&22.11&20.74&19.96&~ \nl
061&0 43 03.5&41 03 35&19.72&18.41&22.36&21.66&20.73&20.00&~ \nl
062&0 43 04.8&41 03 38&$>$20.9&$>$20.2&22.10&21.44&20.76&20.00&~ \nl
063&0 43 04.6&41 03 27&$>$20.9&$>$20.1&21.24&21.92&20.74&19.94&~ \nl
064&0 43 06.7&41 03 30&$>$20.9&$>$20.2&21.38&20.74&20.84&20.04&~ \nl
065&0 43 07.5&41 03 19&20.69&19.70&21.17&20.67&20.84&20.05&~ \nl
066&0 43 07.6&41 02 44&20.33&19.53&20.94&21.20&20.81&20.05&~ \nl
067&0 43 08.0&41 04 43&20.50&19.73&22.11&21.50&20.78&19.97&~ \nl
068&0 43 06.4&41 03 39&20.84&19.94&22.73&22.27&20.80&20.02&~ \nl
069&0 43 08.5&41 03 38&$>$20.9&$>$20.2&21.80&21.09&20.84&20.03&~ \nl
070&0 43 11.2&41 03 34&20.91&19.48&22.09&21.41&20.89&20.13&~ \nl
071&0 43 14.5&41 03 34&$>$21.0&$>$20.2&22.51&21.95&20.88&20.07&~ \nl
072&0 43 10.0&41 06 21&$>$20.9&19.54&22.45&22.04&20.67&19.87&~ \nl
073&0 43 11.1&41 05 11&$>$20.9&$>$20.2&22.46&22.06&20.83&20.06&~ \nl
074&0 43 23.6&41 05 55&19.92&19.29&21.15&20.89&20.90&20.10&eclipsing? \nl
075&0 43 21.6&41 05 03&20.85&19.94&21.91&21.76&20.86&20.09&~ \nl
076&0 43 30.4&41 03 37&20.63&20.00&22.63&22.15&21.11&20.31&~ \nl
077&0 43 17.5&41 05 16&$>$20.9&20.19&22.62&22.09&20.86&20.09&~ \nl
078&0 43 18.0&41 03 59&19.95&19.13&21.93&21.80&20.93&20.15&~ \nl
079&0 43 20.9&41 04 09&20.88&20.07&21.92&21.62&20.90&20.12&~ \nl
080&0 43 16.4&41 05 25&$>$20.9&$>$20.2&22.33&22.14&20.83&20.06&~ \nl
081&0 43 47.0&41 06 05&$>$21.1&$>$20.4&22.33&21.81&21.26&20.44&~ \nl
082&0 43 00.0&41 08 34&20.04&19.58&22.28&22.35&20.23&19.44&~ \nl
083&0 43 02.7&41 09 47&$>$20.6&$>$19.8&22.22&22.55&20.11&19.31&~ \nl
084&0 43 37.5&41 10 09&$>$20.9&$>$20.1&22.45&21.73&20.68&19.95&~ \nl
085&0 43 42.9&41 10 18&$>$20.9&19.88&22.02&21.39&20.81&20.01&~ \nl
086&0 43 45.9&41 10 19&$>$21.0&$>$20.2&22.26&21.78&20.91&20.12&~ \nl
087&0 43 50.1&41 08 37&$>$21.1&$>$20.3&23.22&21.96&21.11&20.27&~ \nl
088&0 43 49.1&41 07 27&$>$21.1&$>$20.4&22.82&22.30&21.19&20.38&~ \nl
089&0 43 37.9&41 10 15&19.42&19.46&22.27&22.27&20.64&19.92&~ \nl
090&0 43 39.6&41 09 06&19.29&19.13&22.06&21.64&20.90&20.07&~ \nl
091&0 43 48.9&41 09 26&$>$21.0&$>$20.3&22.73&22.57&21.03&20.23&~ \nl
092&0 42 45.8&41 11 25&19.65&18.49&21.14&21.23&19.20&18.44&~ \nl
093&0 42 44.3&41 11 29&$>$20.1&$>$19.3&21.88&21.65&19.14&18.31&~ \nl
094&0 42 42.3&41 12 18&17.93&17.65&18.15&17.17&18.73&17.94&nova? \nl
108&0 42 44.5&41 05 54&20.64&19.78&22.39&22.00&20.28&19.44&~ \nl
121&0 42 54.1&41 07 40&$>$20.7&$>$19.9&22.25&22.13&20.28&19.50&~ \nl
140&0 43 04.7&41 14 57&$>$20.0&$>$19.3&22.43&22.65&19.00&18.21&~ \nl
142&0 43 06.7&41 14 31&19.90&19.24&21.38&21.37&19.17&18.36&~ \nl
150&0 43 10.0&41 09 01&20.50&19.44&22.62&22.85&20.38&19.60&~ \nl
156&0 43 14.0&41 09 26&20.54&19.37&21.87&21.49&20.44&19.62&~ \nl
161&0 43 17.5&41 12 12&20.52&19.68&21.55&21.68&20.11&19.34&~ \nl
163&0 43 18.2&41 08 20&20.38&19.83&21.91&21.95&20.52&19.71&~ \nl
165&0 43 20.9&41 10 26&$>$20.7&$>$20.0&21.35&21.89&20.42&19.65&~ \nl
166&0 43 21.6&41 03 25&$>$21.0&$>$20.3&21.21&20.26&20.94&20.17&~ \nl
167&0 43 21.7&41 08 21&20.31&19.55&21.85&21.34&20.50&19.75&~ \nl
170&0 43 23.2&41 10 26&20.14&19.46&21.48&21.55&20.46&19.65&~ \nl
174&0 43 24.7&41 06 11&19.94&19.30&22.46&21.59&20.84&20.04&~ \nl
176&0 43 26.4&41 08 43&$>$20.9&19.57&23.03&21.77&20.66&19.86&~ \nl
177&0 43 26.9&41 09 53&$>$20.8&$>$20.1&22.27&21.73&20.61&19.82&~ \nl
178&0 43 27.3&41 04 04&20.85&19.93&22.82&22.62&21.07&20.29&~ \nl
182&0 43 28.6&41 14 53&$>$20.5&19.56&22.26&21.13&20.02&19.19&~ \nl
184&0 43 29.6&41 17 44&$>$20.4&$>$19.7&21.98&21.08&19.79&19.00&~ \nl
186&0 43 31.9&41 12 19&19.54&19.27&21.50&21.94&20.35&19.59&~ \nl
191&0 43 35.4&41 15 05&20.39&19.72&21.98&22.03&20.28&19.46&~ \nl
193&0 43 35.9&41 09 37&19.11&18.94&21.63&20.89&20.80&20.01&~ \nl
194&0 43 37.1&41 14 31&20.04&19.57&20.57&20.23&20.30&19.48&~ \nl
196&0 43 38.7&41 14 23&$>$20.7&$>$20.0&22.10&21.22&20.44&19.60&~ \nl
197&0 43 39.1&41 12 25&19.24&17.99&22.06&20.96&20.53&19.78&~ \nl
198&0 43 39.5&41 12 35&20.27&19.77&22.29&21.83&20.56&19.72&~ \nl
200&0 43 41.0&41 15 25&20.33&19.64&21.30&21.27&20.36&19.54&~ \nl
203&0 43 44.1&41 10 53&$>$21.0&$>$20.2&22.61&22.36&20.86&20.06&~ \nl
205&0 43 45.7&41 14 52&19.33&18.70&21.73&21.24&20.44&19.67&~ \nl
206&0 43 47.9&41 10 03&19.54&18.92&22.16&21.40&20.91&20.05&~ \nl
207&0 43 48.5&41 09 15&19.57&18.93&21.10&20.59&21.04&20.23&~ \nl
208&0 43 49.9&41 10 44&19.94&19.00&22.28&21.77&20.89&20.09&~ \nl
211&0 43 51.2&41 14 25&$>$20.9&$>$20.1&22.28&21.21&20.66&19.87&~ \nl
302&0 42 37.9&41 07 34&$>$20.5&$>$19.8&22.45&20.49&19.97&19.18&~ \nl
305&0 43 18.1&41 13 23&$>$20.5&19.36&21.73&21.27&19.92&19.11&~ \nl
306&0 43 17.3&41 10 33&20.60&19.36&21.41&21.14&20.38&19.56&~ \nl
307&0 43 21.1&41 09 06&19.99&19.21&21.68&21.33&20.56&19.75&~ \nl
310&0 42 57.3&41 03 45&$>$20.8&$>$20.1&22.77&22.36&20.62&19.87&~ \nl
312&0 42 56.9&41 10 33&$>$20.4&$>$19.7&21.88&21.75&19.86&19.06&~ \nl
315&0 43 14.8&41 11 13&$>$20.6&19.03&21.60&21.50&20.15&19.37&~ \nl
318&0 43 24.7&41 14 10&$>$20.5&$>$19.8&22.24&21.94&19.99&19.21&~ \nl
319&0 43 25.0&41 13 52&$>$20.6&$>$19.8&22.25&22.21&20.08&19.25&~ \nl
320&0 43 25.5&41 13 58&20.12&19.14&21.97&22.12&20.09&19.27&~ \nl
321&0 43 18.5&41 16 29&$>$20.2&$>$19.5&21.96&21.78&19.41&18.60&~ \nl
324&0 43 36.6&41 07 17&$>$21.0&$>$20.2&22.78&22.17&20.99&20.14&~ \nl
325&0 43 54.9&41 08 12&21.13&$>$20.4&21.82&21.38&21.23&20.36&~ \nl
326&0 43 45.5&41 16 16&20.40&19.41&21.22&20.97&20.45&19.64&~ \nl
\enddata
\end{deluxetable}

The search of the KPNO data yielded 139 detected sources over the four
nights of data. Table 1 summarizes the position (columns 2 and 3)
and detected
amplitude range (where the flux difference is expressed as an
equivalent magnitude, columns 6 and 7) 
of these sources in both filters. In addition, we also
list the surface brightness (columns 8 and 9)
and a magnitude at the location of the
variable in the raw reference frame (columns 4 and 5). 
To delineate between variables 
that might be considered resolved in the raw data and 
those that are unresolved we compare the magnitude in the reference frame
at the location of the detected source in the difference 
frame with the local surface brightness fluctuations predicted by 
equation (6) for the best seeing ($1.''1$). 
If the measured magnitude is brighter than twice the predicted 
amplitude of the local surface brightness
fluctuations we quote the measure, otherwise we quote 
a limit of $m_{SBF}-0.75$. Under this criterion, than 
less than half of the sources are resolved at the $2\sigma$ level
in one or both filters. In Figure 7
we show how the detected amplitude range of these sources compares with
the local SBFs in the R filter. Under $1.1''$ seeing only 34 sources have
an amplitude range that exceeds their local SBFs (14 in the I band).
We have therefore achieved sensitivity to both sources and variability
well below anything that can be
considered resolved in the conventional photometry sense.
 
\begin{figure}
\plotone{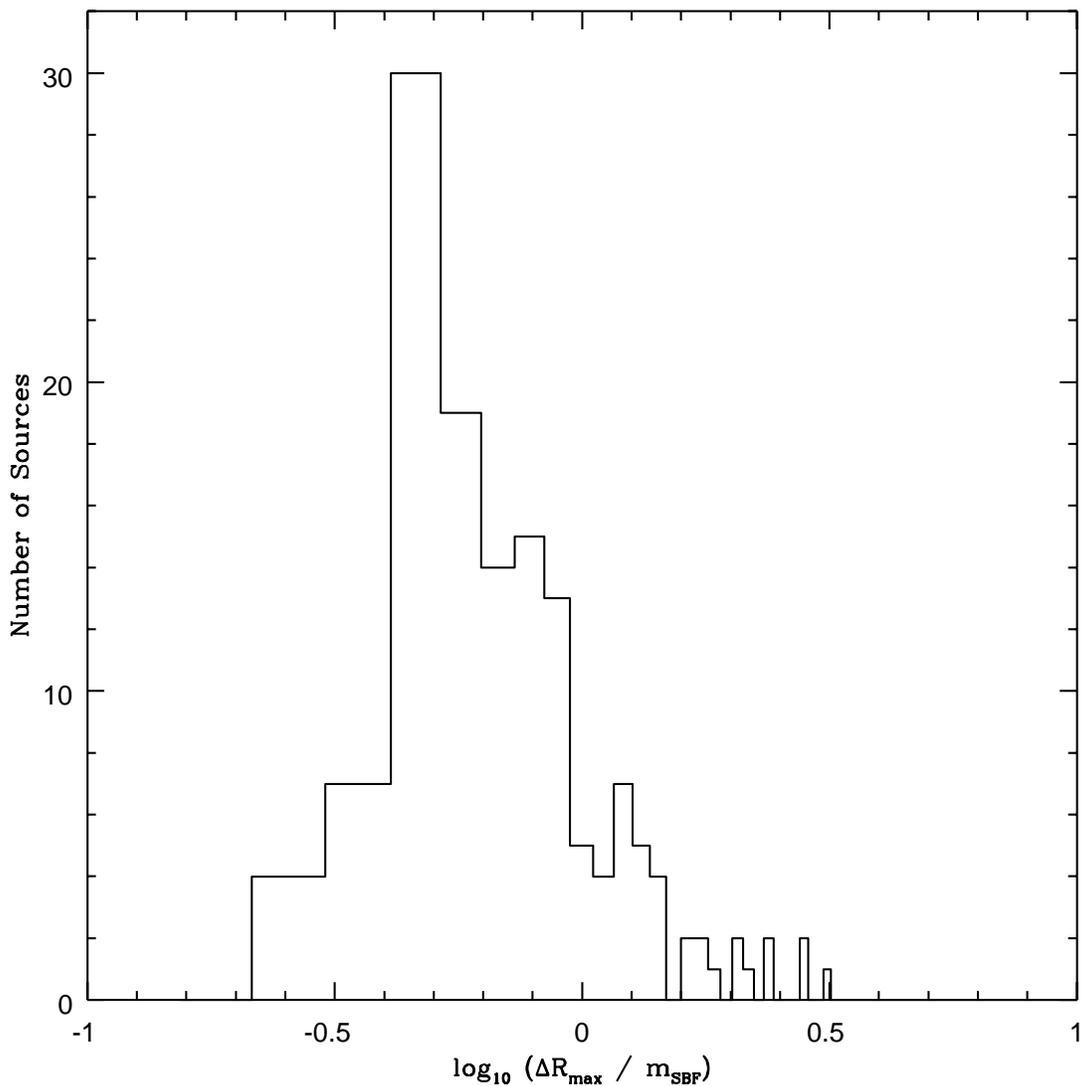}
\caption {
The detected amplitude range in the R band
of the 139 sources discovered in the KPNO data,
expressed as ratio to the statistical amplitude of their local
surface brightness fluctuations under $1''$ seeing (the mean
magnitude of which is $21.^m6$ for these sources).
Under the best seeing of 1.1$''$ only 25\% of the sources
have an amplitude range that exceeds their local
surface brightness amplitude.
\label{fig7}}
\end{figure}

In Figure 8 we show some sample flux difference light curves of objects 
listed in Table 1 that are unresolved in our conservative definition.
All fluxes are relative to the arbitrary reference 
frame of the final night's images. Generally time coverage is too poor
to classify these variables. We expect most of them to
be Classical Cepheids which have periods ranging from 1 to 50$^d$ 
with mean magnitudes of ${\bar {\rm M}}_R = -1$ to -5.5. In addition,
we expect sensitivity to older Pop II Cepheids (W Virginis stars) which
have the same period range, but with ${\bar {\rm M}}_R = -0.5$ to -3.
Other types of variables may include rare RV Tauri stars, and fractional
light curves of long period Mira variables. For recent reviews these
and other variables see Percy (1993) and Nemec \etal (1994).
However, in Figure 9 we present additional flux difference light
curves that we classify as eclipsing variable candidates. These objects
show stability in one or more nights, but a sharp decline in both filters
on one of the two well sampled nights. 

\begin{figure}
\plotfiddle{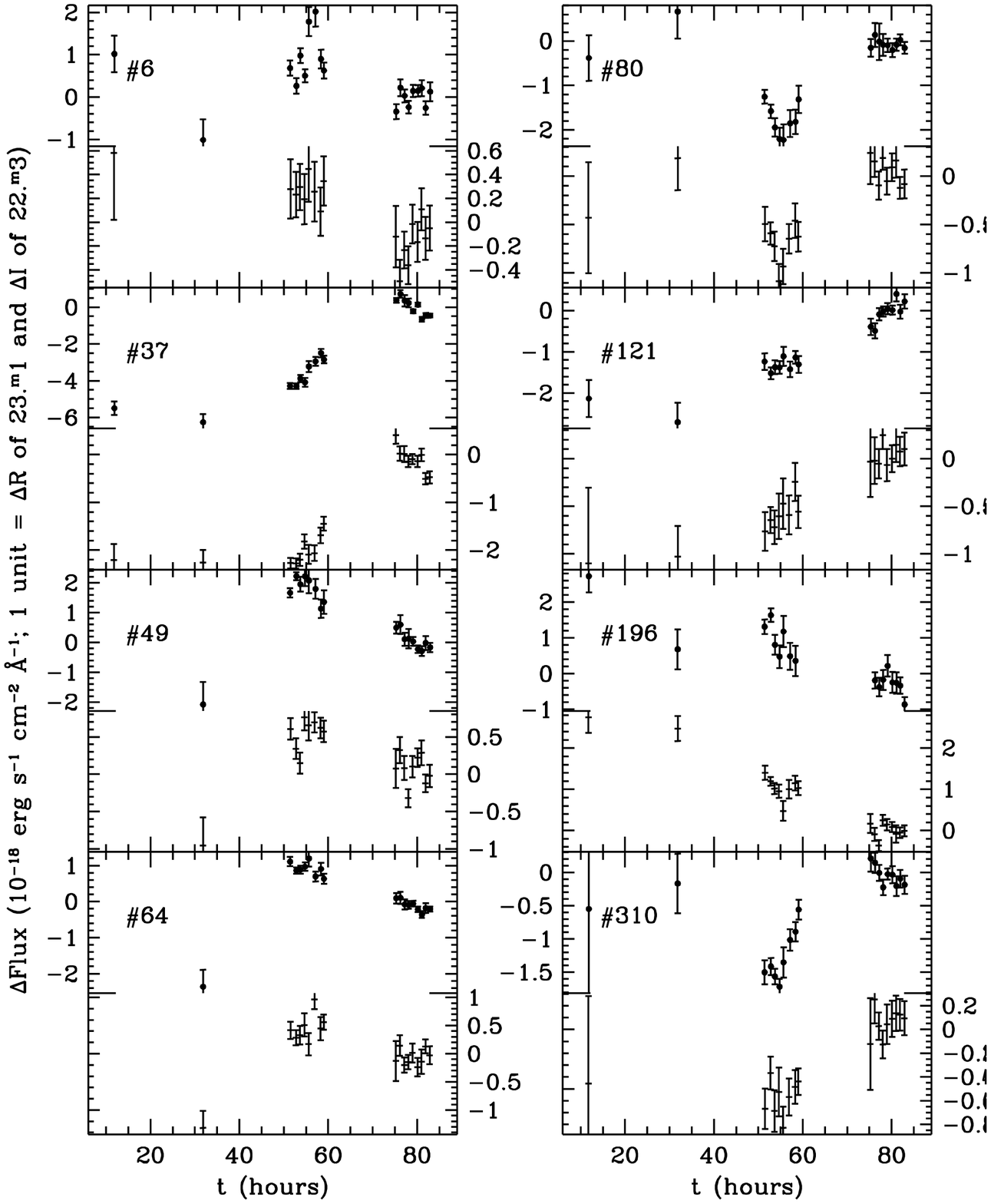}{7in}{0.0}{70}{70}{-216}{0}
\caption {
Sample flux difference light curves in R
and I (upper and lower halves of panels respectively)
of some sources listed in Table 1. The difference fluxes are
relative to an arbitrary mean reference image on final night of the
run. The Julian Date at $t$ = 0 is 2,449,619.50. All these sources
are completely unresolved and showed no evidence in any frame of
flux exceeding the local background by more than $2\sigma$
of the seeing element statistical surface brightness
fluctuations in the original unsubtracted frames.
\label{fig8}}
\end{figure}
 
\begin{figure}
\plotone{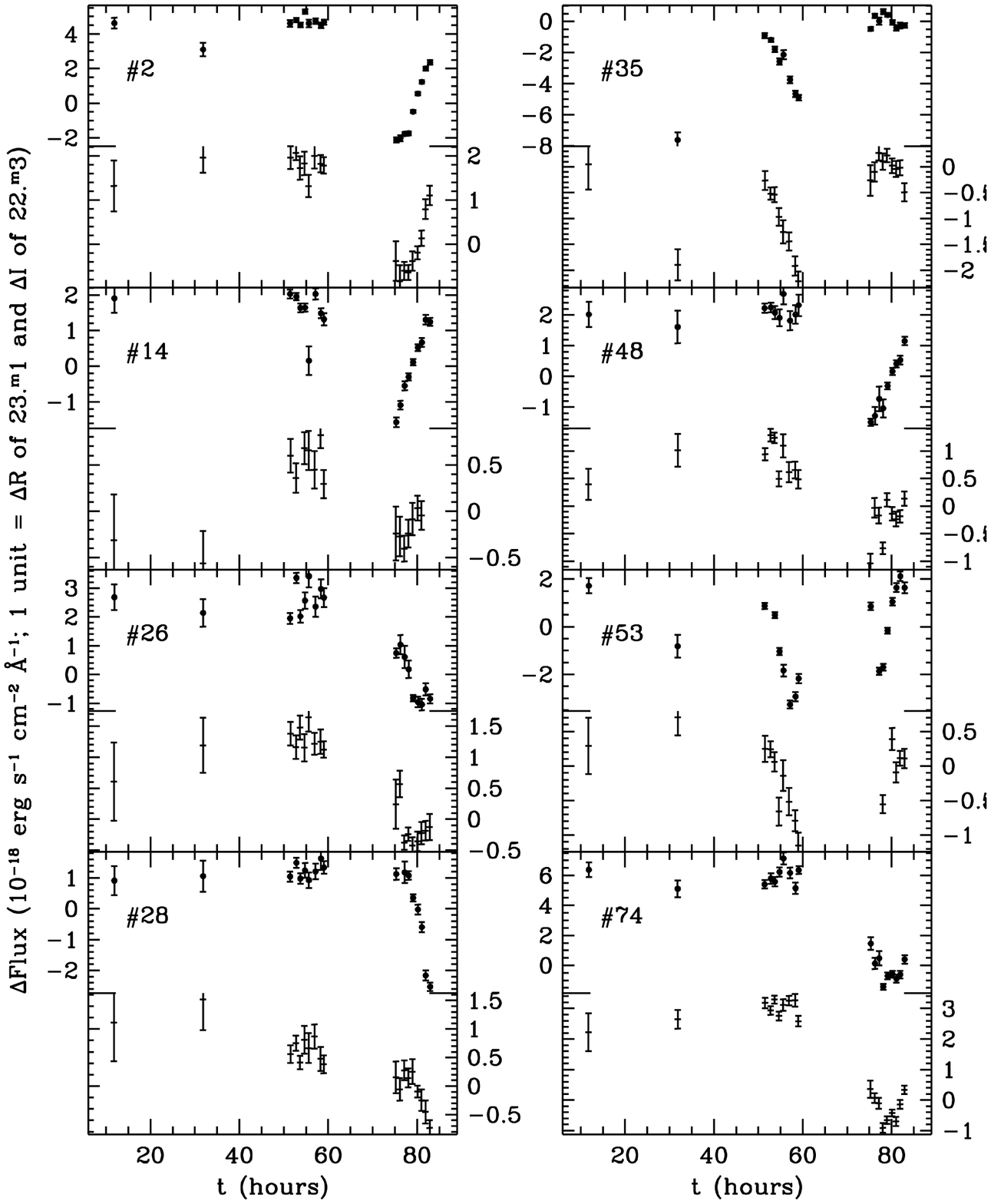}
\caption {
As in Figure 8, but showing some sources in Table 1
that exhibit sharp declines in their flux difference
light curves in both filters,
suggesting that they are eclipsing variables.
The Julian Date at $t$ = 0 is 2,449,619.50.
\label{fig9}}
\end{figure}

We have performed a preliminary analysis of some of the VATT data
({\S 2.3}). In an effort to determine if the same techniques would
recover some of the KPNO variables we have discovered we analyzed
14 consecutive nights of data. Confining the analysis to this
short timespan of the total data sensitizes us to short period
variables that may be similar to the KPNO variables. Of 105 sources
discovered in a search of the difference frames over this period 23 
were found to have a positional coincidence of $< 1''$ to the 85
KPNO sources that fall on the VATT field. These are indicated in 
Table 1. In Figure 10 we compare the KPNO and VATT light curves of 
some of these matches.

\begin{figure}
\plotfiddle{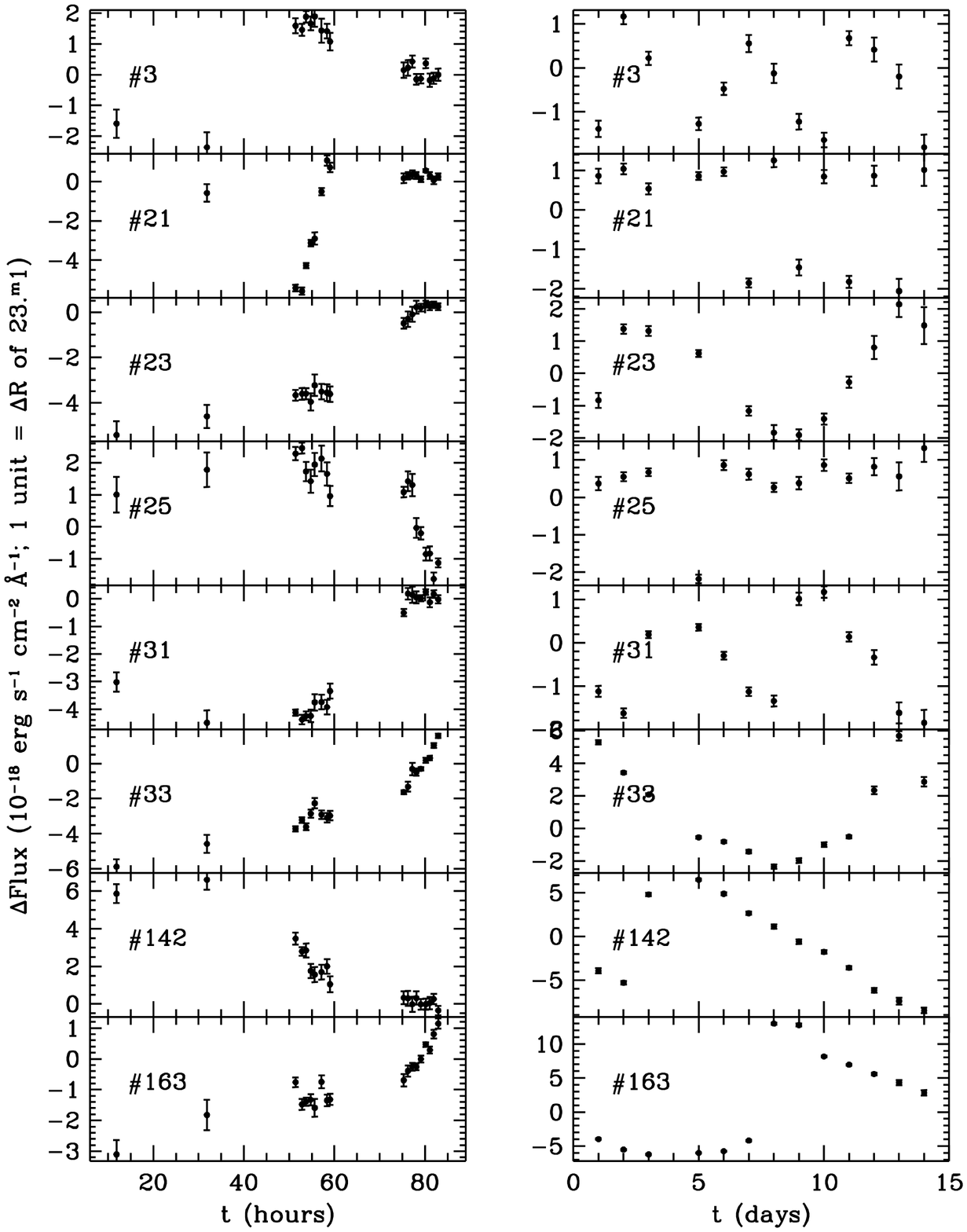}{6.5in}{0.0}{70}{70}{-216}{-36}
\caption {
Comparison of 1994 KPNO (left) and 1995
VATT (right) R band flux difference light curves
for the same sources. The VATT sources were discovered independently
in a restricted 14 night timespan search and found
{\it post facto} to match the position of the KPNO sources.
Notice how in all cases
the behavior of the KPNO light curves is consistent with the
shapes of the VATT light curves which comprise nightly $1-2^h$
exposures. The VATT data also appears to confirm
the eclipsing nature of \#21; its sharp rise, short period
(or half-period) around 2 days.
The Julian Date at $t$ = 0 is 2,449,619.50 for the KPNO light curves
and 2,450,0042.50 for the VATT light curves.
\label{fig10}}
\end{figure}

We expect that a more detailed analysis and search of the data 
(beyond the scope of this paper)
will recover a larger fraction of the KPNO sources. Furthermore, 
we find that differencing VATT images over the timespan of a month 
yields many more variables on the difference frame for our
main far-side field. Figure 11 shows a difference subimage 
($170\times170''$) corresponding to frames taken 40 nights 
apart compared with its original image. Thus we anticipate these data
will generate $> 2000$ light curves. 

\begin{figure}
\plotfiddle{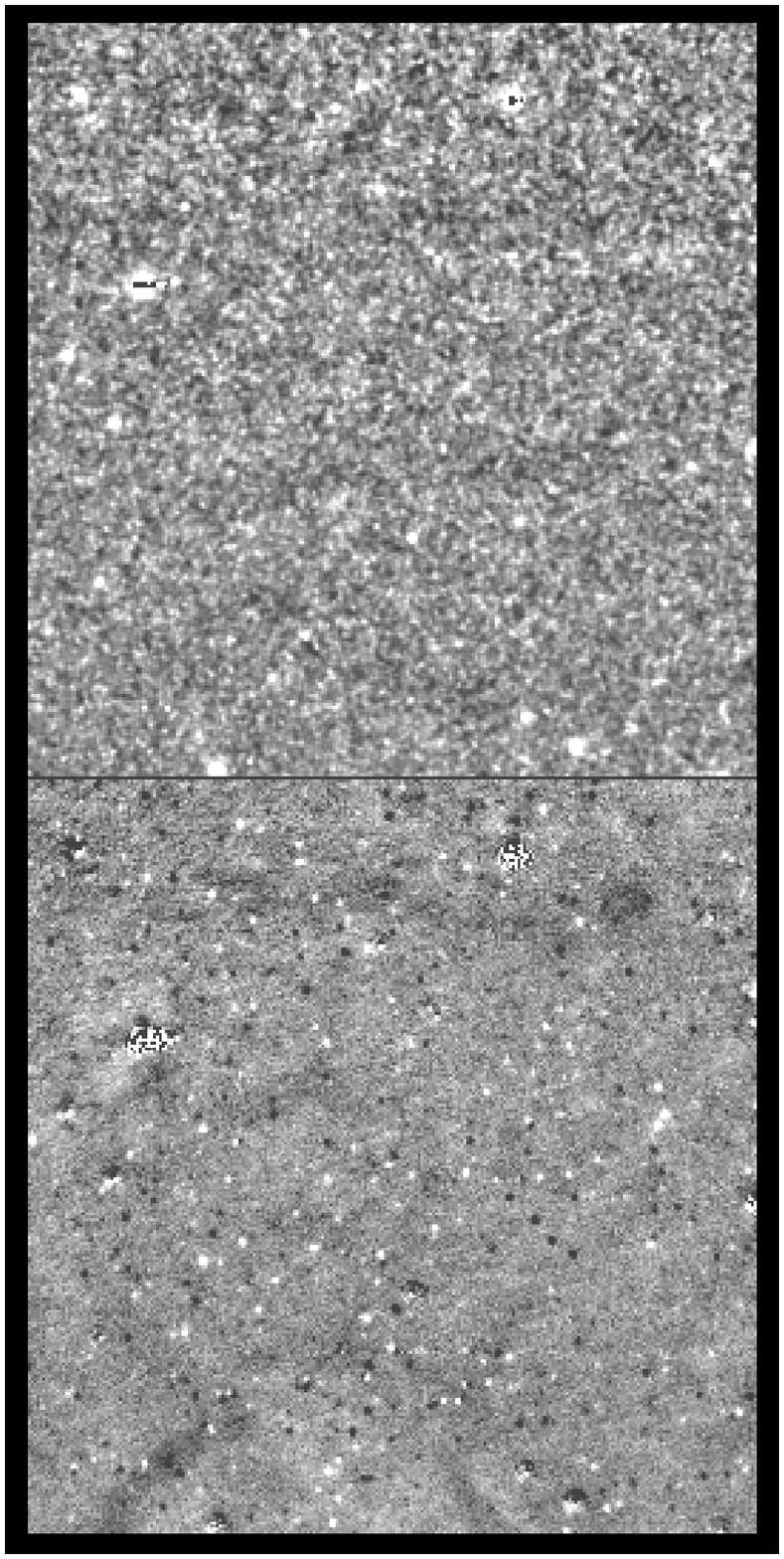}{7in}{0.0}{65}{65}{-288}{-288}
\caption {
An $170\times 170''$ subimage of VATT data taken in
Fall 1995 of the maximal lensing field ({\S 2.2}) on the far side
minor axis of M31.
The top image shows an R band image taken on 1995 October 21
with the a smoothed galaxy background subtracted off to highlight
the unresolved stars,
and the bottom image is the difference image relative to a
similar image taken 38 nights later (shown at 0.2 times the
intensity range of the left image). The difference image shows
many clear point sources in both positive and negative flux, indicating
the detection of variability over this timescale of many stars that
have either brightened or faded. The full difference image contains
$> 2000$ well detected sources.
\label{fig11}}
\end{figure}
 
Interestingly, VATT data indicates a short period (or half-period)
of $< 2^d$ for a KPNO eclipsing variable candidate (\#21 in Figure 10). 
If the object was in M31 its absolute R magnitude
must be $\sim -4$ and with its red color ($R-I > 1.1$) a bright red
giant (M2-M6) would appear to be the only interpretation possible for
identity of the primary star. The minimum size of such as star 
is $\sim 500 {\rm R_\odot}$ (luminosity class $I$ supergiants) and
$\sim 40 {\rm R_\odot}$ (luminosity class $III$, Schmidt-Kaler 1982) 
and the minimum 
period for an object orbiting at its photospheric radius is 
3.5$^y$ and 29$^d$ respectively. 
Thus both the short period and sharpness of the dip 
in the KPNO light curve argue against the object being located 
in M31. However, we note that this object is also difficult to interpret
as a foreground dwarf star system. If we assume the period is $4^d$,
the orbital separation is $\sim 0.05 (M_{tot} / M_\odot)^{-1/3}$ AU,
where $M_{tot}$ the total mass of the system. Assuming equal mass
M-dwarves with $M_{tot} = 1 M_\odot$ their relative velocity is $\sim
140 km/s$. Since the diameter crossing time is $> 8^h$ from the KPNO
light curve, this implies a stellar radius of $5 R_\odot$ - far in 
excess of a typical M-dwarf. Further observations may decide the 
nature of this interesting object.

\section{GALACTIC HALO OPTICAL DEPTH LIMITS}

\subsection{Stellar Number Densities}
 
The number of stars per pixel detectable above a certain $S/N$ threshold is
crucial in understanding the conversion of an event rate to an optical depth
due to gravitational lensing.
In turn, the number of stars per pixel depends on the shape of the luminosity
function, and its first-moment integral, the surface brightness.
The surface brightness determines not only how many stars brighter than a
certain magnitude are present in a pixel, but also the background flux which
sets the $S/N$.

Calibrating our surface brightness data is straightforward, despite the lack of
a night sky brightness determination, in that surface brightness photometry is
published for our field (Walterbos \& Kennicutt 1987).
The flux/ADU/pixel and night sky brightness (assumed to be constant over our
sequence-averaged frame) is simply recovered from comparison of the slope and
zero-intercept of surface brightness and counts across the frame.
 
We have developed a simple technique for recovering the shape of the 
luminosity
function, based on the distribution of local surface brightness in various
pixel.
Figure 12 shows the number of pixels in a small subframe of our image 
plotted as a histogram versus the signal per pixel in ADU, once the 
sky background has been subtracted.
The thick solid line shows the plot for our actual data, for a subframe in
which the spatial gradient of counts has been removed without affecting the
mean count per pixel.
The other curves denote pixel histograms for several cases of simulated star
fields composed from luminosity functions of the form $\phi(L) dL \propto
L^{-0.4\alpha} dL \propto 10^{\alpha m} dm$, where $m$ is apparent magnitude
(in which $\alpha=0.60$ for the thin solid line, 0.55 for the dotted.)
The normalization of the simulated histograms is set only by requiring 
that the first moment of the distribution, the surface brightness, is 
maintained.
We note that higher moments are also recovered simultaneously, allowing the
entire shape of the real data's histogram to be recovered with the proper
choice of $\alpha$, in this case $\alpha = 0.59 \pm 0.01$.
This value of $\alpha$ is similar to those found for bright stars in
the Solar Neighborhood, $\phi$, (Luyten 1968), in the outer disk of M31 (Hodge,
Lee \& Mateo 1988), or in K band, for the inner disk of M31
(Rich, Mould \& Graham 1993).
These simulated images were constructed using stars spread over 
$17 < R_j < 28$
according to the above power laws.
On the faint end we used a Solar Neighborhood (Luyten 1968) $\phi$. The
two distributions on the bright end were joined by requiring 
continuity at $R = 28$. 
At magnitudes brighter than $R=17$ negligibly few stars appear in the
simulations, while fainter than $R=28$, 
stars are too uniformly spread to affect
the count fluctuations per pixel by more than 2\% of the total.
The power law approximation must break down at the faint end of this
distribution, however, given the presence of the horizontal branch at
$R=25.0$.
To reach this level in measuring the luminosity function, 
an instrument such as $Hubble$
$Space$ $Telescope$ is required.
Our technique of simulated pixel histograms could then be employed with the
$HST$ data to reach below the horizontal branch, where the luminosity function
is probably once again well-approximated by a power law (or some other
few-parameter family).
 
\begin{figure}
\plotone{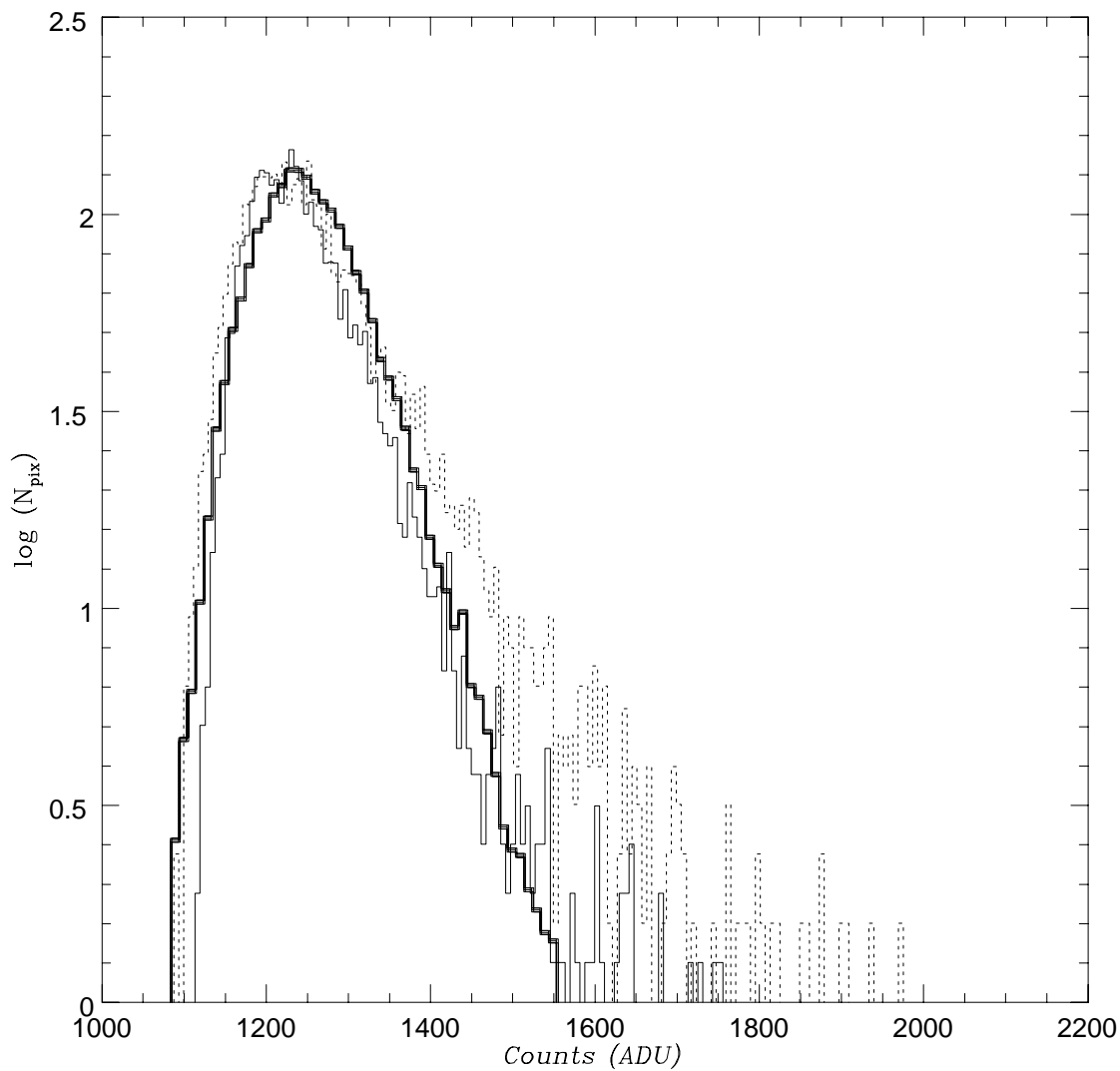}
\caption {
A histogram of residual pixel intensities of a $150\times150$ pixel region of
KPNO data ({\S 2.2}) after the underlying galaxy
gradient has been subtracted off (bold line).
The skewness of the pixel distribution
is primarily determined by the ratio of bright to faint stars in the
luminosity function. A simulated image of artificial stars distributed with
a power law luminosity function (see {\S 6.1}) gives a good fit
with $\alpha = 0.59$ to the pixel histogram.
\label{fig12}}
\end{figure}
 
Knowing the luminosity function, we can estimate the number of detectable 
stars in the sample.
From our best LF fit we estimate at this location there
are 0.35 per square arcsecond above R = 22.5. This compares to an
estimate of 0.4 stars per square arcsecond scaling the Hodge LF measured
in the disk of M31 (Hodge, Lee \& Mateo 1987) 
using the count/brightness ratio
({\it c.f.,} C92), the latter estimate being
affected disproportionally by a large disk component.
Therefore we consider our estimate to be more realistic and
adopt it for the purposes of this paper. Thus for our integrated galaxy
surface brightness of R = 4.67 we estimate there are 
$6.7\times10^5$ stars on our
field exceeding R = 22.5.
 
The usual technique for calculating event rates consists of choosing a 
stellar sample above a threshold apparent brightness, and a minimum
amplification, which implies a certain lensing cross-section per 
MACHO given a known Observer-Lens-Source (O-L-S) geometry.
The product of the implied optical depth corresponding to the number 
density of MACHOs and the number of stars in the sample produces the
mean number of lensing events at a given time, and the number 
of events over a period of observation (much longer than the event) is
then inversely proportional to the typical event duration (Griest 1991).

In a sample of lensed stars surveyed by the DIP 
technique, however, we cannot impose either stellar brightness nor
amplification thresholds, but instead a flux cutoff $f_{min}$. The
number of detectable events, therefore, is computed by considering 
which minimum amplification (Paczynski 1986):

$$ A_{min} = {u_{min}^2 + 2 \over {u_{min}(u_{min}^2+4)^{1/2}}}, \eqno(7)$$

\noindent
corresponding to an impact parameter $r_{min}$ at the point of 
maximum amplification ($u_{min} = r_{min} / R_e$, where $R_e$ is
the projected Einstein Radius [{\S 6.2}]), implying 
a lensing cross-section $\sigma = \pi r_{min}^2$ for a 
detectable event. $A_{min}$ depends on the brightness of the star,
$f_*$, and $f_{min}$: $A_{min} = f_{min} / f_* + 1$, and the number
of events at a given time, 

$$ N_{ev} =  \int^\infty_0
{ N_*(f_*) \tau (A_{min}) df_*}$$

$$ ~ = a_1 \int^\infty_0
{ \phi_* (f_*) \sigma (A_{min}}) df_*, \eqno(8)$$

\noindent
where $a$ absorbs multiplicative factors relating $N_*$ to $\phi$
(such as the field of view) and $\tau$ to $\sigma$ (such as O-L-S 
geometry and the number density of MACHOs). 
Comparing this integral to the analogous expression in the case of
thresholds in $A$ (for the sake of discussion, a 34\% enhancement, as 
for $u=1$) and $f_*$:

$$ N'_{ev} = a_1 \int^\infty_{f_{min}} { \phi_* (f_*)
(\pi R_e^2) df_*}. \eqno(9)$$

\noindent
Regardless of $\phi$, $N_{ev}' < N_{ev}$. For the case of the 
Solar Neighborhood luminosity function (Luyten 1968), 
$N_{ev} /  N_{ev}' \approx 3$, with most of the additional events
due to high amplifications of stars with $f_* < f_{min}$.
In the case of a large horizontal branch contribution to $\phi$,
however, the number of high$-A$, small $f_*$ events could be 
even larger.

Due to these additional events, the distribution of amplification 
of detected sources is {\it not} described by a uniform distribution
in impact parameters, as is found for the $f_{min}$ $A-$threshold case
characteristic of microlensing searches using resolved stars. In the 
DIP case, there are more low$-u$, high$-A$ events, depending on the
shape of $\phi$. Given sufficient S/N, however, we can measure $A$
from the shape of the light curve (even though we don't know $f_*$),
meaning that $\phi$ can be recovered from the distribution in $A$.
(Although at very high $A$, the dependence of the shape of the light
curve on $A$ becomes degenerate.)
In {\S 6.3} we show that high S/N has been achieved. In a future
paper we hope to apply the technique to actual microlensing 
events. 

\subsection{Mass and Timescale Sensitivities}
 
Before turning to the individual surveys themselves we must examine
the masses of MACHOs to which we are sensitive to and the event timescales
such masses correspond to. The projected Einstein radius of a MACHO
of mass $M$
at the lensed source at some observer-source distance $D_{os}$ is
given by,
 
$$ R_e^2 = {4GM\over c^2} {D_{ol} D_{ls} \over D_{os}}, \eqno (10) $$
 
\noindent
(Paczynski 1986) where the $D$'s are the indicated O-L-S distances.  
A detectable event corresponds to when this projected radius 
exceeds radius of the lensed star (below this limit the light 
is simply being redistributed within the image of the star's photosphere).  
Let us consider a survey sensitive to stars in M31 brighter 
than R = 22.5, including high amplification events of fainter stars.
We adopt our mean lensing source to be a K0 III star with a mean radius 
corresponding to $\sim 12 {\rm R_\odot}$ (Schmidt-Kaler 1982). 
Consider a MACHO in M31's halo with $D_{ls} = 10~kpc$.
This corresponds to a minimum MACHO of mass $\sim 5\times10^{-5}{\rm M_\odot}$,
but many of these stars are larger, so thereby produce a partial cutoff at
larger MACHO masses.
For a Galactic MACHO, however, we have the benefit of the observer-lens
proximity, in which the effects at the lens plane are projected to the source
plane, resulting in a much smaller mass for a given angular size.
The minimum mass here for $D_{ol} = 10~kpc$ corresponds to $7\times10^{-9}
{\rm M_\odot}$ ($\sim$0.002 Earth masses).
 
Now consider the timescales corresponding to these mass limits.
Since the stars in the inner disk of M31 rotate at 260~km~s$^{-1}$ (Braun
1991), and objects in the inner halo/outer bulge move transversely at an
average 230~km~s$^{-1}$, the typical timescale $t_{_E}$ for minimal mass
M31 MACHO/M31 star events is 6.5 hours.
The component of the Sun's motion transverse to the M31 sightline of
170~km~s$^{-1}$, and the typical Galactic MACHO transverse velocity of
180~km~s$^{-1}$ result in a typical timescale for minimal Galactic MACHO events
of 8 minutes.
 
\subsection{Optical Depth Calculation}

None of the light
curves of the objects in Table 1 are appear consistent with 
microlensing events on the $< 8^h$ timescale. Furthermore no
frame showed single ``spikes'' in both filters that might be
due to a high amplification lensing event on a timescale of
$< 50^m$ which is the time resolution of the individual difference
frames in this analysis. 
A more sophisticated and complete search of
this dataset as well as an analysis of frames at the full time
resolution of the individual images ($10^m$) using the search 
technique proposed by Gould (1996) may yet reveal 
interesting candidates, but such a effort is beyond the 
main purpose of this techniques paper. However, based on our
simple search we can still make an estimate of some interesting 
optical depth limits. We note that these
estimates are largely limited by our uncertainty of the luminosity
function.

Sources were originally identified by eye in the difference frame at
typical S/N ratio of 6 or greater. Thus these S/N ratios correspond to 
$12\sigma_{photon}$ given our determined technique limits in 
{\S 4.1}. For a fiducial minimum amplification of $>34\%$
corresponding to stars passing within the Einstein ring of the lens
(Paczynski 1986), we must therefore detect the original star 
at $36\sigma_{photon}$ to detect this minimum amplification at a S/N
$> 6$ in the difference frame. Given our estimate of the luminosity
function and the number of stars we are sensitive to
on the single coadded frames in the analysis in {\S 6.1}, we estimate
that each difference frame is sensitive to $> 34\%$ amplifications
of $2.0\times 10^5$ stars ($R > 21.2$). However, we are also sensitive
to a larger number of fainter stars that suffer higher amplifications
({\S 6.1}) giving us sensitivity to $6.1 \times 10^5$ stars.

Requiring multiple exposures per event in the KPNO data, in both bands, we have
19 independent 50$^m$ sample times, corresponding to a mass scale of about
$2\times 10^{-7} {\rm M}_\odot$.
Eliminating ``edge'' points over the four nights gives 13 independent sample 
times (confined to the two good nights). 
Thus at this timescale we are sensitive to $7.9\times10^6$
star-epochs, which corresponds to a $2\sigma$ optical depth of $\tau
< 5\times 10^{-7}$.
(The minimal $\tau$ grows by $[\rm M/(2\times 10^{-7} {\rm M}_\odot)]^{1/2}$.)
Over 8$^h$ timescales corresponding to 
masses of $8\times 10^{-5} {\rm M}_\odot$ we have two sample times
corresponding to the two last nights, which corresponds to a 
$2\sigma$ limit of $\tau < 3.3\times 10^{-6}$ 
applying both to M31 and Galactic halo MACHOs.
These limits are expressed in terms of a delta-function concentration of MACHO
mass at the value of M discussed.
Given Paczynski's (1986) estimate for a simple spherical halo of
the optical depth of the Galactic halo towards M31 of 
$\tau = 1.0\times 10^{-6}$, we can conclude that in two nights we have 
eliminated the possibility at the 2-sigma confidence quoted
above that the Galactic halo is comprised of a single mass population
of MACHOs in the Earth mass range (${\rm M}_\oplus = 
3\times 10^{-6} {\rm M}_\odot$)
({\it c.f.}, the comparable EROS Galactic halo 
limits of Aubourg \etal 1995 and MACHO results of Alcock \etal 1996
in this range).
For M31 plus Galactic MACHOs, we can expect values for $\tau$ of 5-10$\times
10^{-6}$ (C92, Han \& Gould 1996), which is a factor of several times larger
than our 2-sigma limit quoted above for $8\times 10^{-5} {\rm M}_\odot$.
Despite the systematic uncertainties involved, it would appear that these mass
ranges are ruled out for a 100\% contribution to the both dark matter MACHO 
halos.

Another possible source of microlensing in our field is {\it intergalactic}
masses, as might be associated with dark matter clustering with
galaxies, but not actually resident in their halos.
We show that we are not currently sufficiently sensitive to detect these.
First, if these were non-baryonic masses (e.g.~primordial black holes: Crawford
\& Schramm 1982), they might be able to contribute up to $\Omega_{DM} \approx
1$ to the Universe's mean density (being irrelevant to Big Bang nucleosynthesis
constraints on baryonic matter density).
Second, if they fall into potential wells as cold dark matter, and dominate the
matter distribution, their density between the Galaxy and M31 might be an order
of magnitude greater than the universal critcal density $\rho_{_{crit}}$. 
(Models
studying the Local Group in a CDM universe are currently being produced:
Governato 1996.)
If one assumes a model whereby the region beyond the galaxy halos are dominated
by a uniform distribution at $\rho_{_{crit}}$ of such objects, their
contribution to the optical depth would be about $1.5\times 10^{-8}~h^2$ 
($h ==
H_o/100$ km s$^{-1}$).
Even a local enhancement of ten would be undetectable in this survey,
but within the reach of surveys we are preparing for the near
future.
Intergalactic microlensing events involving M31 giant and supergiant stars as
sources would be limited to masses larger than about $10^{-5} M_\odot$.
Such events would have a signature of a significantly longer timescale than
those of similar masses in either halo.
Their projected cross-section per unit mass is much larger given their more
favorable placement roughly half-way between the observer and source
(see equation
~10), and their transverse velocities with respect to the sightline are
probably lower since they are not within the gravitational potential of a
galaxy.
These factors, accounting for their distance from the source plane, combine to
make these events about four times longer than Galactic or M31 halo events due
to the same lens mass.

\section{DISCUSSION and CONCLUSIONS}

We have demonstrated that detecting variability among completely
unresolved stars is a tractable problem through an image differencing
technique that 
involves registering, photometrically scaling and PSF-matching
pairs of frames. From this approach we have derived optical depth limits 
for MACHOs in both the Galactic and M31 halos ({\S 6.3}).
The technique is not limited by 
seeing variations or stellar density. 
We have shown how, in a typical wide field imager, PSF matching a pair
of frames is complicated by small changes in the telescope focus,
and developed an algorithm which uses a limited number of stars to
model the full frame matching PSF function. This algorithm has general
applicability to wide-field imagers at different telescopes
and differencing images taken through separate systems. We have 
shown how we have achieved residuals in difference frames taken 
at the KPNO 4-m prime-focus camera that are within a factor of three
of the predicted seeing element photon noise. These difference
frames taken over four nights yielded 139 variable sources most of
which evidenced no flux at any stage more than twice the local
surface brightness fluctuations and thus be considered completely 
unresolved and beyond the capabilities of conventional 
crowded field photometry. A preliminary analysis of a sample of 
data taken at the VATT 1.8-m telescope one year later and using 
the same techniques outlined in this paper has confirmed 
the reality of a large fraction of these KPNO sources.  
These VATT data were taken using an optical corrector which facilitates
the DIP process by requiring only a few (1-4) PSF adjustments to 
match frames.

This techniques paper is primarily motivated by the potentially 
rewarding aspects of a microlensing survey of M31 suggested by 
Crotts (1992). However, the technique has wide applicability to
microlensing (see Gould 1996) and more general variability surveys
(see also Phillips \& Davis 1995). We have attempted to provide
in {\S 4} an assessment, 
both qualitatively and quantitatively, of all systematic effects 
pertinent to our KPNO survey and our ongoing VATT survey. We recommend
Gould's analytical treatment of the DIP approach 
for a more thorough quantitative assessment as applied
to more general microlensing surveys. 

Other microlensing surveys directed to
the Galactic Bulge and the LMC can apply the DIP
technique to their datasets at hand. In no longer
being limited by sensitivity to only well detected and
resolved stars, a DIP
analysis will dramatically increase the sensitivity to microlensing
events. In particular, however, such an analysis will also be 
sensitive to short timescale events (limited only by the exposure times 
and time resolution of their data) associated with very 
short timescales, especially very high amplification 
``caustic-crossing'' lensing events associated with binary/multiple
component lenses and, perhaps, other optical transients ({\it c.f.}, 
Hudec \& Soldan 1994).

We also note that despite the PSF undersampled images in
the present WFPC2 camera onboard {\it HST} and the consequences
that this has to image differencing technique ({\S 4.6} and 
Gould 1996), the algorithm pioneered by Mighell (Mighell \& Rich 1995)
to cope with the unsampled PSF and substantially improve the photometric 
accuracy over a standard photometric software such as DAOPHOT 
and DOPHOT may well also have an application as an avenue 
by which HST data maybe exploited for pixel analysis ({\it e.g.},
Gould's [1995] microlensing survey of M87).

Finally, we note that, in principle, the techniques we have 
outlined are applicable not only to searches for 
photometric variability, but also {\it astrometric variability}.
Changes that we can detect at approximately the photon shot-noise level can
also be caused by simple relative motion of the sources, not just flux changes.
In this regard DIP is also a sensitive tool for studying proper motions
at the milli-arcsecond level,
since the maximum change in flux, occuring at radius $= \pm \sigma/\sqrt{2}$,
(where $\sigma$ is the r.m.s. width of the image), grows as $\Delta f/f =
0.72 \Delta r / r_{HW}$, where $\Delta r$ is the change in the star position,
$r_{HW}$ is the HWHM radius for a Moffat seeing function, and $\Delta f$
the absolute value of the change in flux which vaies from positive to
negative values in the direction of motion. Thus for a star detected at
$100\sigma$ in $1''$ seeing, even in crowded conditions, motions as small
as 14 mas can be detected at the 1$\sigma$ level.

Since a requirement of image differencing is accurate registration of all
stars on a frame, it is worth noting
that a pixel analysis of colour shifted events (Kamionkowski 1995)
might be very revealing, since a difference image is sensitive not
only the flux changes but {\it positional shifts}. We might expect a
colour shifted event to show such a shift since the ratio of light
of the lens and lensed star changes. The difference image quantifies
precisely the systematics involved in such a problem, since the
residuals around all other stars in the frame are a measure of
how good the registration between frames is as well as how the PSF
has been matched. The residual image position should rest at the location
of the component being lensed within the blend.
Over longer timescales one might also think of
measuring proper motions of stars in a the Galactic Bulge and
LMC by differencing images taken a few years apart
- any shift in a star's position will be revealed by the residuals
in the difference frame. Of course, if a proper motion could be
measured this way then most stars on the frame will show ``proper
motion'' residuals, but for examination of a known lensing event
there maybe some gain of information of how the lensed star is moving
relative to most stars in the field.

\acknowledgments
We would like to express our sincere thanks to Dave DeYoung (Director,
KPNO) for the allocation of Director's 
Discretionary Time on the 4-m telescope
for this project. We are grateful to Andrew Phillips (Lick) for providing
us with his PSF matching software. We also thank the Vatican Observatory
Research Group, in particular Richard Boyle and Chris Corbally in 
assisting our efforts on the VATT, and the Steward Observatory engineering
group, in particular Richard Cromwell, Dave Harvey and Steve West. 
We would also like to express thanks to Andrew Gould for useful 
discussions and suggestions during the writing of this paper. 
This research was funded by the David and Lucile Packard Foundation.

\clearpage

\end{document}